\documentclass[3p]{elsart}
\usepackage{graphicx}
\usepackage{multirow}
\usepackage{amssymb}
\usepackage{amsmath}
\usepackage[switch]{lineno}

\newcommand{\ZHAireS}{\mbox{ZHA\scriptsize{${\textrm{IRE}}$}\normalsize{\hspace{.05em}S}}\hspace{.45em}}

\begin{document}

\begin{frontmatter}

\title{Monte Carlo simulations of radio pulses in atmospheric showers using ZHAireS}

\author[USC]{Jaime Alvarez-Mu\~niz,}
\author[USC]{Washington R. Carvalho Jr.,}
\author[USC]{Enrique Zas}

\address[USC]{Depto. de F\'\i sica de Part\'\i culas
\& Instituto Galego de F\'\i sica de Altas Enerx\'\i as,
Universidade de Santiago de Compostela, 15782 Santiago
de Compostela, Spain}

\begin{abstract}

We present predictions for the radio pulses emitted by extensive 
air showers using ZHAireS, an AIRES-based Monte Carlo code that takes
into account the full complexity of ultra-high energy cosmic-ray induced
shower development in the atmosphere, and allows the calculation of the
electric field in both the time and frequency domains. We do not presuppose 
any emission mechanism and our results are compatible with a superposition of
geomagnetic and charge excess radio emission effects. We investigate the
polarization of the electric field as well as the effects of the refractive index $n$ and shower geometry on the radio pulses. We show that geometry, coupled
to the relativistic effects that appear when using a realistic
refractive index $n>1$, play a prominent role on the radio emission of air showers.  

\end{abstract}

\end{frontmatter}

\section{Introduction}
In the last decades, the study of ultra high energy cosmic rays (UHECR) has
been one of the most active areas in astroparticle physics~\cite{springeruhecrreview}. The detection of
extensive air showers (EAS) created by UHECR in the atmosphere has been
accomplished mainly with two detection methods. The first consists on detecting the particles of
the cascade reaching the ground using an array of particle detectors. 
The second method uses telescopes to detect the fluorescence photons emitted by the
particles in the shower as they travel through the atmosphere. Only the latter
method is able to directly measure the EAS longitudinal development in the
atmosphere, but is subject to a very low duty cycle ($ \sim10\%$),
since it can only be used in clear moonless nights. The surface array
and fluorescence techniques are used simultaneously at
the Pierre Auger cosmic ray observatory~\cite{auger}.

Radio emission from EAS was first observed by Jelley et al.~\cite{jelley} in 1964.  Theoretical and experimental
research in this area was very active in the
sixties and early seventies. We refer the reader to a review by
Allan~\cite{allan} and references therein. Nevertheless, interest in this
technique declined in the late seventies, mainly due to radio
interference~\cite{lopes}. 
Recent developments in high speed electronics and information 
technology have renewed interest in the detection of radio emission in the MHz
range from the particles in EAS. The radio 
detection technique is, in principle, sensitive to the longitudinal development of
the shower, like the Fluorescence technique, but its duty
cycle is much higher, since it could be used anytime
except during thunderstorms. 

This resurgence of the radio technique has taken form as new experiments have
been developed,
such as CODALEMA~\cite{codalema}, LOPES~\cite{lopes,lopesgeo,lopesldf}
and AERA~\cite{aera}, accompanied by new calculations of the
radio emission in EAS, which include analytical techniques with different levels of
sophistication~\cite{Huege_analytical,Scholten_MGMR,Montanet_radio,Ardouin_Coulomb}, Monte Carlo
methods~\cite{ReAires,Konstantinov,REAS2,REAS3,SEIFAS,zhaires_air11} and semi-analytical methods~\cite{Scholten_MGMR_sims}.

In this work we present ZHAireS~\cite{zhaires10,zhaires_air11} (ZHS+AIRES), a new simulation of radio
emission in EAS that combines the full shower simulation capabilities
of AIRES~\cite{aires} with specific algorithms developed to calculate
the electric field emitted by particles in dense media showers, implemented in the well tested ZHS 
code~\cite{ZHS92,ARZ10}. These algorithms are obtained from first principles, so no
emission mechanism or model is presupposed. The radio emission calculations are done
in parallel to the AIRES shower simulation: As each charged particle in
the shower is propagated by AIRES in steps, each propagation step is taken as a single
particle track and its contribution to the radio emission is calculated and
added to the total electric field in both the time and frequency domains. This
procedure naturally accounts for interference effects associated to the
different space-time positions of the particles in the shower. 

Other Monte Carlo simulations of radio emission of air showers exist, such as
REAS~\cite{REAS2,REAS3}, where the electron and positron tracks from CORSIKA~\cite{corsika}
simulations are first histogrammed and then used to generate random $e^{\pm}$
trajectories, which in turn are used for the radio emission calculations in
the time domain only. An earlier version of this code, REAS2~\cite{REAS2}, was based
only on geosynchrotron emission, but was shown to be inconsistent with later
simulations~\cite{REAS3,zhaires_air11}, which in turn are based on a pretty
generic algorithm~\cite{allan} that has been used for a long time to
simulate pulses generated in dense media~\cite{ZHS92,ARZ10}. Another difference is that ZHAireS uses a model for the
variation of the refractive index with altitude, while REAS uses a fixed
refractive index equal to unity in its calculations.

This paper is organized as follows: In section \ref{sec:emission} we give a
short overview of the main radio emission mechanisms in air showers and the
specific formalism used for emission calculations in ZHAireS, including a
model for the variation of the refractive index with altitude. In section
\ref{sec:1Dmodel} we develop a one-dimensional toy model useful for
understanding the main characteristics of the emission and stress the
prominent role played by geometry. In section
\ref{sec:results} we show results of
ZHAireS simulations, discussing the influence of the
distance from the antenna to the shower core, the refractive index and the
shower zenith angle on the electric field pulse. We also discuss the spectrum
of the radio emission. In section
\ref{sec:polarization} we analyze the polarization of the electric field and
discuss how it can be used to separate the contributions from  various
emission mechanisms. Finally, in section
\ref{sec:conclusion} we conclude the paper.

\section {Radio Emission in Air Showers}
\label{sec:emission}
\subsection{Main emission mechanisms}
\label{sec:mechanisms}
In air showers, the dominant mechanism responsible for the radio emission is
believed to be the deflection by the geomagnetic field of electrons and
positrons in the shower~\cite{codalemageo,lopesgeo,falckenature}. Several
approaches have been developed to model this emission. Since the
direction of the Lorentz force depends on charge, it
leads to a spatial separation of electrons and positrons in the shower, that
can be thought of as a moving macroscopic dipole and a transverse
current traveling 
through the atmosphere at a speed $v \approx c$ along with the shower
front, and which has been the basis for macroscopic 
calculations~\cite{kahnlerchegeo,Scholten_MGMR,Scholten_MGMR_sims}.
 Another approach,
which is adopted in this work, is to calculate the emission for each
particle trajectory, i.e. a microscopic
approach~\cite{ZHS92,ARZ10,zhaires10,zhaires_air11,REAS3}.

The geomagnetic mechanism has a quite clear signature. Its polarization is
anti-parallel to the direction of the Lorentz force, i.e. in the direction of
$-\vec{\beta}\times\vec{B}$, where
$\vec{\beta}$ is the speed of the particle in $c$ units and $\vec{B}$ 
the geomagnetic field~\cite{kahnlerchegeo,lerchenature,codalemavxb,scholtenarena2010}.

Another emission mechanism also thought to be important in EAS is the Askaryan
effect. It was first proposed
by Askaryan~\cite{Askaryan62}, who suggested that coherent radiation
could be emitted by showers in which a charge excess develops. This mechanism, which
dominates the radio emission of showers in dense media, is also known as the charge excess mechanism. 
For a particle following a straight track at constant speed, it can be shown \cite{Jackson}
that:
\begin{equation}
\vec{E}_{rad}\propto -e [ \hat{u} \times ( \hat{u} \times \vec{\beta} ) ]
\end{equation}

\noindent So the electric field emitted by  particles traveling along the
shower axis lies in the plane defined by the direction of the shower axis
($\vec{\beta}$) and the observation direction $\hat{u}$. Furthermore, it is
perpendicular to $\hat{u}$ and its direction depends on the charge of the
particle, pointing towards (away from) the shower axis for negative (positive)
charges \cite{ZHS92,ARZ10}. This means that if there is no charge excess in the shower, the net
electric field is zero. However it is well known that knock-on interactions
and Compton scattering incorporate electrons from molecules of the medium into the shower, leading to an
excess of moving negative charges \cite{Askaryan62}, which in turn are responsible for a net electric field
with a radial polarization w.r.t. the shower core. 

\subsection{Radio emission calculation in the \ZHAireS code}
\label{newcode}

\begin{figure}
\begin{center}
\scalebox{0.8}{
\includegraphics{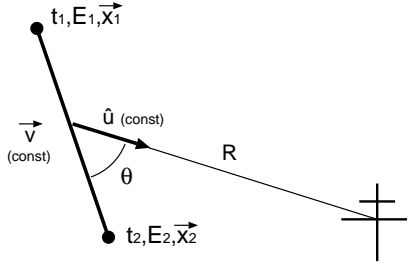}}
\caption{Scheme of a single particle track. Speed and direction to the
  observer are assumed constant inside each track.}
\label{fig:singletrack}
\end{center}
\end{figure}

The algorithms used for the calculation of radio emission in ZHAireS are based
on ZHS algorithms~\cite{ZHS92,ARZ10}. These were derived from first
principles, namely from the Lienard-Wiechert potentials, and thus do not assume any emission mechanism. The full derivation can be seen in~\cite{ZHS92} for frequency domain calculations, and~\cite{ARZ10}
for the time domain. The trajectories of shower particles are divided in 
tracks, which can be made arbitrarily small. Both the speed of the
particle and the direction to the observer are assumed not to vary 
over the track. In Fig. \ref{fig:singletrack} we show a schematic picture of such a
track. A convenient division of trajectories in small tracks is already
performed by Monte Carlo simulations of shower development in
order to propagate the particles in the shower. The ZHS
algorithm is then used to calculate the contribution of each track to the
net emission of the shower, accounting for any interference between
tracks. This approach automatically takes into account the contribution to the electric field due to the start,
end, and any change in the direction and energy of each particle track. Thus
any kind of deflection, scattering, creation or annihilation of a charged
particle, due to any physical process used to simulate the shower is taken
into account in the radio emission\cite{zhaires_air11,zhaires10,endpoint}.

In the original ZHS algorithms~\cite{ZHS92,ARZ10}, 
the field calculation is performed in the Fraunhofer approximation. 
In this case the observer ``sees'' the radiation
arriving from all the points in the shower at the same angle, and only changes
in the phase of the contribution from each point are taken into account through a
simple projection. This approach works well for
showers in dense media, such as ice, since the size of the shower is much
smaller than the typical distance to the antennas. But in the case of air
showers, the distance to the observer is usually of the order of the shower size, and
thus the Fraunhofer approximation breaks down. In order to make the method
valid for the closer observers in air showers, we allow the
distance and direction to the observer to change from one track to the next. The direction and distance to the
antenna are still taken to be constant inside a single track (see
Fig.~\ref{fig:singletrack}), i.e. we still use the
Fraunhofer approximation inside each track. It can be
shown~\cite{inpreparation} that this procedure reproduces the expected behavior
$E\propto 1/\sqrt{R}$ of the field~\cite{hunearfield}, where $R$ is the
distance between the charge and the observer. Since we still use the Fraunhofer approximation inside a single track of
length L, the following condition has to be satisfied:

\begin{equation}
\frac{L^2\sin^2{\theta}}{R}<\frac{\lambda}{2\pi}\;\;
\label{eq:fraunhoffer}
\end{equation}

\noindent where $\lambda$ is the wavelength of the emission. In the
simulation, the vast majority of tracks satisfy this condition for frequencies
up to 300 MHz, which is much higher than the frequency of the maximum of the
emission, $\sim 1-30$ MHz. If a particular track in the simulation does
not satisfy this condition, e.g. a track passes very close to an antenna, it is
further divided into several sub-tracks, and the field calculation is performed using
a different R and $\theta$ for each sub-track.

The positions $\vec{x}_{1,2}$, times $t_{1,2}$ and
kinetic energies $E_{1,2}$ for the beginning and end points of the track are
obtained directly from AIRES. This, along with the position $\vec{x}_{ant}$ of
the antenna is all that is needed for the calculation of the contribution to
the vector potential $\vec{A}(t,\hat{u})$ due to a given track:
\begin{equation}
\vec{A}(t,\hat{u})=\frac{\mu e}{4\pi R
  c}\vec{\beta}_\perp\frac{\Theta(t-t^{det}_1)-\Theta(t-t^{det}_2)}{1-n\vec{\beta}\cdot\hat{u}}
\label{eq:timefresnel}
\end{equation}  
\noindent where $\vec{u}=R\hat{u}$  is the vector from the middle point of the
track to the antenna, $\vec{\beta}=\vec{v}/c$, $\vec{\beta}_\perp=-\left[\hat{u}\times(\hat{u}\times
  \vec{\beta}) \right]$ is the projection of $\vec{\beta}$ onto a plane
perpendicular to $\hat{u}$, 
$t^{det}_{1,2}=t_{1,2}+nR/c-n\vec{\beta}\cdot\hat{u}~(t_{1,2}-t_0)$ are the
retarded (detection) times for the beginning and end of the track,
respectively, $t_0=(t_1+t_2)/2$ is the average time for the track and
$\Theta(x)$ is the Heaviside step function\footnote{In the formalism, as
  described in ~\cite{ARZ10}, the limit of eq. (\ref{eq:timefresnel}) for
  $(1-n\vec{\beta}\cdot\hat{u})\rightarrow 0$ is
  used  instead of eq. (\ref{eq:timefresnel}) if the denominator vanishes.}. Note that Eq.~(\ref{eq:timefresnel})
is written in the radiation gauge, since we disregard the static term of the
Lienard-Wiechert potentials in its derivation~\cite{ARZ10}. 

To obtain the radio signal at each antenna for the
shower as a whole, the contributions of each particle track to
the vector potential $\vec{A}(t)$ are added up\footnote{Note that
  interference effects are taken into account automatically, since the
  contribution of each track to the vector potential is related to a specific retarded
  time.}. This net vector potential is then differentiated with respect to
time to obtain the net electric field as a function of time for each
antenna. In the far field, this formalism is equivalent to the one used
in~\cite{REAS3}, as shown in~\cite{endpoint}. 

Besides the field calculations in the time domain, ZHAireS can calculate the
radio emission in the frequency domain as well. The radiation term of the electric field is $\vec{E}(t)=-\partial \vec{A}/
\partial t$. Applying this to the vector potential expression (eq. \ref{eq:timefresnel}) we obtain~\cite{ARZ10}:
\begin{equation}
\vec{E}(t,\hat{u})=-\frac{\mu e}{4\pi R
  c}\vec{\beta}_\perp\frac{\delta(t-t^{det}_1)-\delta(t-t^{det}_2)}{1-n\vec{\beta}\cdot\hat{u}}
\label{eq:timefresnel-EF}
\end{equation}

In the ZHS formalism~\cite{ZHS92,ARZ10}, we use the following convention for the
Fourier transform:
\begin{equation}
\tilde{f}(\omega)=2\int_{-\infty}^{\infty}{f(t) e^{i\omega t}dt}
\label{eq:ft}
\end{equation}
Applying this Fourier transform convention to Eq.~(\ref{eq:timefresnel-EF}) we obtain~\cite{ZHS92}:
\begin{equation}
\vec{E}(\omega,\hat{u})=-\frac{\mu e}{2\pi R
  c}\vec{\beta}_\perp\frac{e^{i\omega(t-t^{det}_1)}-e^{i\omega(t-t^{det}_2)}}{1-n\vec{\beta}\cdot\hat{u}}
\label{eq:freqfresnel}
\end{equation}
\noindent which is the expression we use for the Fresnel regime
frequency-domain calculations in ZHAireS. The contributions of each track to
$\vec{E}(\omega)$ are added up to obtain the net spectrum at each antenna.

\subsection{Variable refractive index}
\label{sec:n(h)}

As stated before, in ZHAireS we use a model for the variation of the
refractive index $n$ with altitude in the atmosphere. In fact there is a strong dependence of $n$
on temperature, pressure and humidity. To take the dependence with altitude 
into account we use an exponential model for the
variation of the refractivity ${\cal R}$ with height $h$,
motivated by the exponential decrease of density with altitude,  
\begin{equation}
{\cal R}(h)={\cal R}_s \exp{(-K_r h)}
\label{eq:n(h)}
\end{equation}
where ${\cal R}(h)=[n(h)-1]\times 10^6$ is the refractivity at an altitude
$h$ in km and we used ${\cal R}_s={\cal R}(h=0)=325$ and
$K_r=0.1218~{\rm km^{-1}}$. These values reproduce the refractivity calculated
in~\cite{Gerson} up to $h\sim10$ km within less than $\sim 1\%$. 
The values in \cite{Gerson} take into account the humidity dependence of the
refractivity,  which increases the refractivity at low altitudes where
the effect of a variable index of refraction 
is most important, since the shower has a larger number of
particles. At higher altitudes, the exponential model slightly
overestimates the refractivity, but the effect of the variable index of 
refraction at these altitudes is much less important, since above 20
km the shower has barely started developing and the number of particles is small. 
It is also worth noting that since ${\cal R}$ depends on temperature
and humidity, there are large seasonal and even daily variations in
the refractivity that are associated to atmospheric conditions and cannot thus be
accurately described by any model\footnote{If the humidity as a function of
height can be monitored using e.g. a LIDAR, then more accurate values of the
refractivity for specific atmospheric conditions could be calculated
by adding a humidity term to Eq. (\ref{eq:n(h)}), as described in~\cite{Gerson}.}.  

The variable atmospheric refractive index would in principle make the radio
emission follow a curved path. However, in \cite{Scholten_MGMR_sims} it was shown that the effect of
the deviation from a straight path on the time structure of the pulse is
negligible, and thus in ZHAireS we assume that the radio emission follows a
straight path from the emission point to the antenna.
However, we explicitely take into account the effect of a variable refractive index in the propagation time of the signal. For this purpose we calculate an effective refractive index 
$n_{eff}$ for each particle track in the ZHAireS shower simulation:
\begin{equation}
n_{eff}=1+{\cal R}_{eff}\times10^{-6},\;\;\;{\cal R}_{eff}=\frac{1}{R}\int_{0}^{R}{{\cal R}(h)~dl}
\label{eq:neff}
\end{equation}
\noindent where $R$ is the distance from the track to the observer, and
$dl$ is an infinitesimal length along that path whose altitude $h$ varies along 
the path. It is important to note that while $n_{eff}$ is used for the
calculation of the retarded times, $n(h)$ is used for the angular dependence of the
emission, given by the denominator of  Eqs. (\ref{eq:timefresnel}) and
(\ref{eq:freqfresnel}), i.e. the Cherenkov angle depends only on the
refractive index at the emission altitude.

We expect the effect of the variable refractive index to be more important
at relatively low altitudes, where $n$ is larger.
Since $n$ in our model varies between $n\sim 1$ at high altitudes and
$n\sim1.000325$ close to ground, we expect the effect of the variable $n$ on
the pulse characteristics discussed in section
\ref{sec:1Dmodel} (such as start-time, duration and compression in time) to be
between the effects obtained for $n=1$ and $n=1.0003$. This can be clearly
seen in Figs.~\ref{fig:tdetxdepth} and \ref{fig:fc-varn} below.

\section{One-dimensional toy model}
\label{sec:1Dmodel}
We can gain much understanding of the main features of the 
radio pulses in the time-domain with the aid of a very simple 
model. We show in this section that many of the characteristics
of the pulse such as start-time, peak value of the electric field and duration in time,
are mainly determined by the geometry of the system formed by the shower 
and the observer. In the model 
we assume a vertical atmospheric shower in which particles propagate 
along the shower axis at the speed of light $c$, and we ignore their lateral spread. 
For the moment we assume a constant index of refraction $n$.
We do not make any assumption on the radio emission mechanism, 
and we just account for the retarded time, i.e. the radio signal is emitted
at a time $t'$ and reaches the observer located at a distance $r$ to the 
shower core at a later time $t=t'+\Delta t_p$, where $\Delta t_p$ is given
by the electromagnetic wave travel time between the emission point and
the observer. The model can also be applied to understand the
radio emission properties in dense media~\cite{ARZaskaryandensemedia}.

\begin{figure}
\begin{center}
\scalebox{0.55}{
\includegraphics{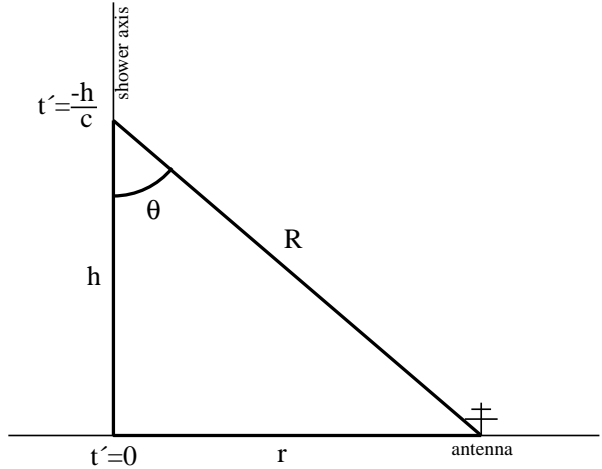}}
\caption{Sketch of the one-dimensional model. The shower development is represented
by a 1D line neglecting the lateral structure. 
The shower front is assumed to travel at a speed $c$ and reach ground at a time $t'=0$.
Radio emission from a height $h$ travels along a distance $R$ before
arriving at the observer located at a distance $r$ to the shower axis. }
\label{fig:geotoy}
\end{center}
\end{figure}

In Fig.~\ref{fig:geotoy} we show a sketch of the model.
The origin of time $t'=0$ is arbitrarily fixed at the time 
at which the shower front reaches ground.
The emission time at a height $h$ above the ground is $t'=-h/c$,
the propagation time to an antenna on the ground at a
distance $r$ from the shower core is $nR/c$, 
where $R=\sqrt{h^2+r^2}$, and hence the arrival time of
the radio emission at the antenna is:

\begin{equation}
 t=\frac{n \sqrt {h^2 + r^2}-h }{c}
\label{eq:tdet}
\end{equation}

The 1D model although simple is useful to understand at least
in a qualitative way the prominent role played by the geometry 
in the behavior of the electric pulse signal. 

\subsection{Start-time of the electric field pulse}

The start-time time of the electric field pulse is given by  
the minimum value of $t$. By doing $\partial t/\partial h =0$,
one can find the height of the shower seen first by the observer: 

\begin{equation}
h_{start}=\frac{r}{\sqrt{n^2-1}}
\label{eq:htmin}
\end{equation}

Substituting $h_{start}$ in Eq.~(\ref{eq:tdet}), we obtain the
time at which the observer sees the onset of the radio pulse:

\begin{equation}
t_{start}=\frac{r}{c}\sqrt{n^2 -1}
\label{eq:tdetmin}
\end{equation}

For $n>1$, $t_{start}$ is linear in $r$ and the observer sees an increasing delay in the start time of the pulse 
as the distance to the shower core increases. 

\begin{figure}
\begin{center}
\scalebox{0.65}{
\includegraphics{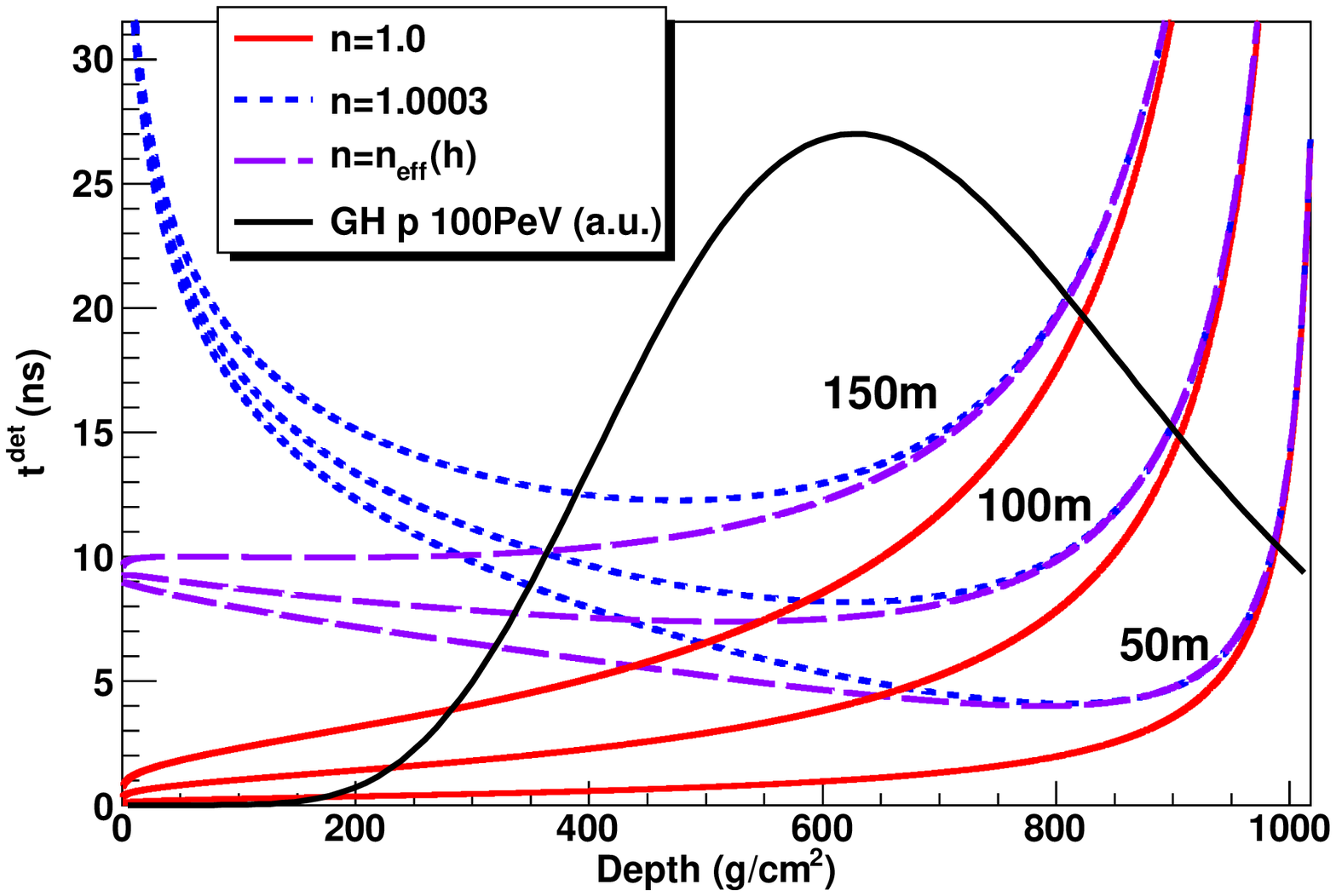} 
}
\scalebox{0.68}{
\includegraphics{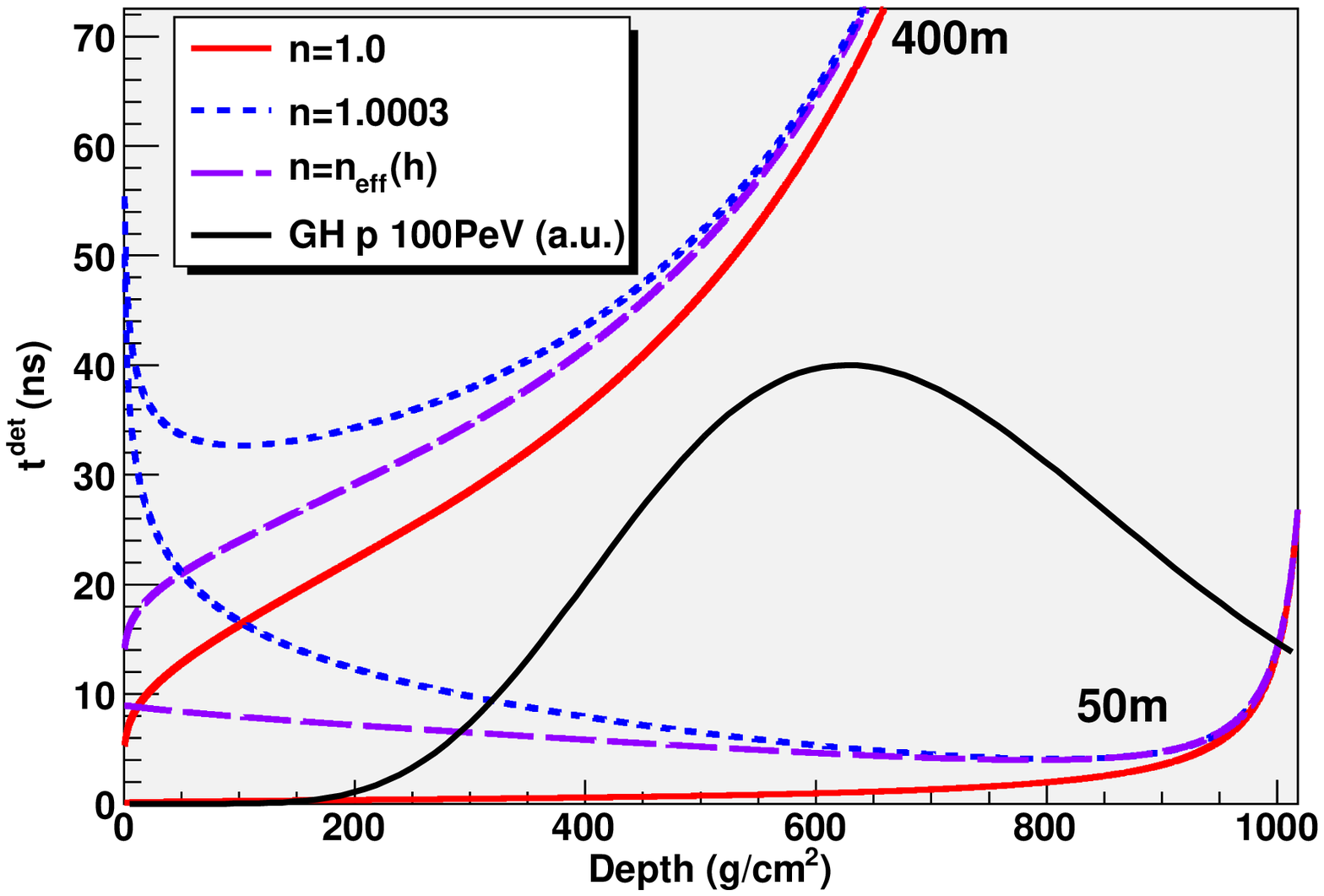}
}
\caption{Relation in Eq.~(\ref{eq:tdet}) between the observer time $t$ and the 
time at the source $t'$ or equivalently as depicted in the plot the depth of shower 
development. The relation is valid for vertical showers and is shown for observers 
at different distances $r$ to the shower 
core and for different values of the index of refraction $n$. Also shown is the longitudinal 
profile of a 100 PeV proton shower (from a Gaisser-Hillas function) in arbitrary
units, which gives an idea of the relative number of particles at different depths.}
\label{fig:tdetxdepth}
\end{center}
\end{figure}

In Fig.~\ref{fig:tdetxdepth} we plot the relation given in Eq.~(\ref{eq:tdet}) between 
the observer time $t$ and the time at source $t'$ or equivalently 
the shower depth (see also \cite{{Scholten_MGMR_sims}}). The relation is shown at different distances $r$ to the
shower core and assuming different values of $n$ including the more realistic 
$n$ varying with height (see section \ref{sec:n(h)}). One can see the increase of $t_{start}$ with $r$ when 
$n>1$ as predicted by Eq.~(\ref{eq:tdetmin}). Also, the angle $\theta$ between the observer and the shower axis (see Fig.~\ref{fig:geotoy}) is given by
$\tan\theta = r/h$. When $h=h_{start}$ as given in Eq.~(\ref{eq:htmin}) 
it is straightforward to show that: 

\begin{equation}
\tan\theta_{start}=\sqrt{n^2-1}=\tan\theta_C
\label{eq:theta_start}
\end{equation}
and the observer sees first a shower height $h_{start}$ at time
$t_{start}$ with an angle equal to the Cherenkov angle. A similar analysis
with similar results to the ones described above was developed independently
and in parallel with ours by de Vries, Scholten and Werner~\cite{scholtenprl},
and in dense media in~\cite{ARZ11}.

Due to relativistic effects associated to the speed of the shower
(assumed to be $c$) being larger than the speed of the radio waves ($c/n$),
at a fixed detection time $t_d$ an observer at a
distance $r$ from the shower core
may see two different stages of shower development (two different heights $h_\pm$) simultaneously. This can be seen in Fig.~\ref{fig:tdetxdepth}. Solving the quadratic Eq.~(\ref{eq:tdet}) for $h$ with $t_d$ constant we obtain:

\begin{equation}
h_\pm = \frac{ct_d \pm nc \sqrt{t_d^2-r^2(n^2-1)/c^2}}{(n^2-1)}
\end{equation}

Two different real solutions exist if $n>1$ and the argument 
of the square root is positive, which is equivalent to the condition
$t_d>t_{start}$. This apparent violation of causality is simply a relativistic
effect due to the particles traveling faster that the speed of the
pulse in the medium. When $n=1$ the shower is seen by
any antenna away from the core in a ``causal'' way, from beginning to
end, as can be seen in Fig.~\ref{fig:tdetxdepth}, provided the observer
is not located on the shower axis (when $n=1$ and $r=0$ the whole
shower is seen at the same instant of time). 

When $r$ is sufficiently large or $n=1$ (see Eq. (\ref{eq:htmin})), $h_{start}$
can become larger than the actual height at which the shower starts
developing, namely the height $h_0$ where the first interaction
occurs. The emission from
heights above the first interaction point is due to the primary particle
only and can be neglected. In that case, the observer would start to see
the onset of the pulse at a time corresponding to the height $h_0$: 

\begin{equation}
 t_0=\frac{n \sqrt {h_0^2 + r^2}-h_0 }{c}
\label{eq:t0}
\end{equation}

The height $h_0$ can be considered a physical limit on $h_{start}$, which
corresponds to the depth  $X_0$ of the first interaction. The
corresponding start times $t_0$ can be read from Fig.~\ref{fig:tdetxdepth} by
truncating the curves in Fig.~\ref{fig:tdetxdepth} at the minimum depth $X_0$. 
So, for any given $n$, there is a critical distance $r_{crit}$, at which
$h_{start}=h_0$:

\begin{equation}
 r_{crit}=h_0\sqrt{n^2-1}
\label{eq:rcrit}
\end{equation} 

For $n>1$, as we move away from the shower axis the start time first increases
linearly with $r$ (following Eq.~(\ref{eq:tdetmin})) until $r=r_{crit}$. For
larger $r$ it follows the non-linear behavior of Eq.~(\ref{eq:t0}). For
$n=1$, the effective $t_{start}$ is always given by $t_0$ 
in Eq. (\ref{eq:t0}), regardless of the distance $r$ to the core. 

Regarding the relativistic effects described above, when $n=1$,  $r_{crit}=0$
and the shower is seen in a causal way by any observer, as stated before. For $n>1$, observers at
distances greater than $r_{crit}$  will also see the shower in a ``causal''
way, starting at $h_0$ , and in this model they never see the shower at the
Cherenkov angle.

\subsection{Peak value of electric field pulse}
\label{sec:1Dmodel-peak}

The peak value of the radio pulse is mainly determined by three factors,
namely, the distance from the shower to the observer, the number of charged
particles in the shower, and also and in a very important way by geometrical
effects associated to a time compression factor. The relation
between the observer time and the shower height, given by Eq.~(\ref{eq:tdet})
and shown in Fig.~\ref{fig:tdetxdepth}, is non linear. One can consider a
given observer time interval and obtain from this relation the interval of the
shower development that contributes to it. Clearly, the portion of shower
development that contributes will be large when the slope of this relation is
small. We can define a compression factor $f_c$ taking the derivative of Eq.~(\ref{eq:tdet}) with respect to $h$: 
\begin{equation}
f_c=\left|\frac{\partial t}{\partial h}\right| = \left|\frac{1}{c} \left(1-\frac{nh}{\sqrt{r^2+h^2}}\right)\right|~~~{\rm [ns~m^{-1}]}
\label{eq:compression}
\end{equation} 
The absolute value accounts for the fact that the derivative of $t$ with respect to $h$
changes sign when $h=h_{start}$ (i.e. $t=t_{start}$), corresponding to
a reversal in the time sequence of the shower as seen by the observer. 
The inverse of the compression (or Doppler) factor $f_c$ can be thought of as
a measure of the compression in time~\cite{Gousset_inclined}.  Qualitatively, a small value of $f_c$ implies that the emission from a
relatively large portion of the shower contributes to the pulse in
a relatively small interval of observer's time $\delta t$, increasing
the pulse with respect to other cases in which the factor $f_c$ is larger.
Moreover, when $h\rightarrow h_{start}$ (i.e. $t\rightarrow t_{start}$) then
$f_c\rightarrow 0$ and $\delta t\rightarrow 0$. This is related to the fact
that the observer sees the shower at $t=t_{start}$ with an angle equal to the
Cherenkov angle, as shown in Eq.~(\ref{eq:theta_start}). In fact, since
$\cos\theta=h/R$ it is straightforward to show that the factor in
Eq.~(\ref{eq:compression}) is proportional to $|(1-n\beta\cos\theta)|$. As a
consequence of $f_c\rightarrow 0$, the factor 
$(1-n\beta\cos{\theta})$ in the denominator of Eq.~(\ref{eq:timefresnel}) goes
to zero, tending to enhance the peak value of the pulse, seen at the
Cherenkov angle\footnote{This singularity is not a problem for our numerical
  calculations for several reasons: For the singularity
to actually enter the numerical calculation, $f_c$ would have to vanish
(within the precision of our code) at precisely the middle of the track (see
Fig.~\ref{fig:singletrack}). This (almost) never happens, but if it does, the
code uses the limit of Eq.~(\ref{eq:timefresnel}) for $f_c\rightarrow0$
\cite{ARZ10}. Also note that just like in reality (and Monte Carlo too) there
is a finite time resolution (we use $\delta t = 0.5$ ns in this
work), which spreads very high (and short) vector potential peaks over the whole time bin.} 

It is important to note that although a small $f_c$ induces a very big
effect in the pulse in the 1D model where all particles follow the
shower axis, in a real shower this effect will be less
important due to the lateral and angular spread of the particles in the shower. 
The lateral and angular spread will change the observer angle and increase
the compression factor at $h_{start}$ with respect to what the 1D model
predicts ($f_c\sim 0$), since $f_c\propto
-\cos{\theta}$. Furthermore, the lateral spread of the particles in a real
shower will also change the distance to the observer, smearing the arrival
time of the signal.

\begin{figure}
\begin{center}
\scalebox{0.35}{
\includegraphics{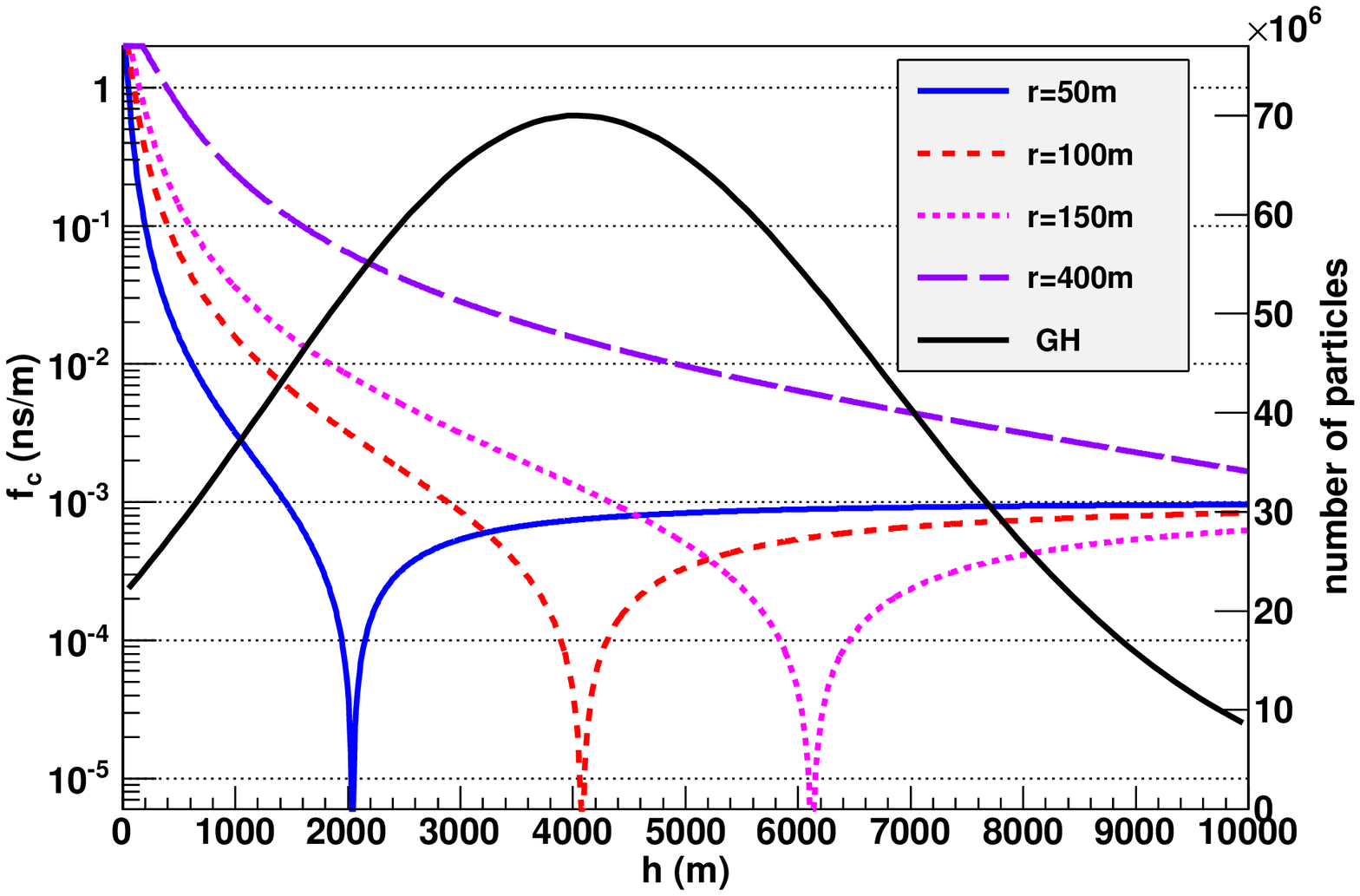}\includegraphics{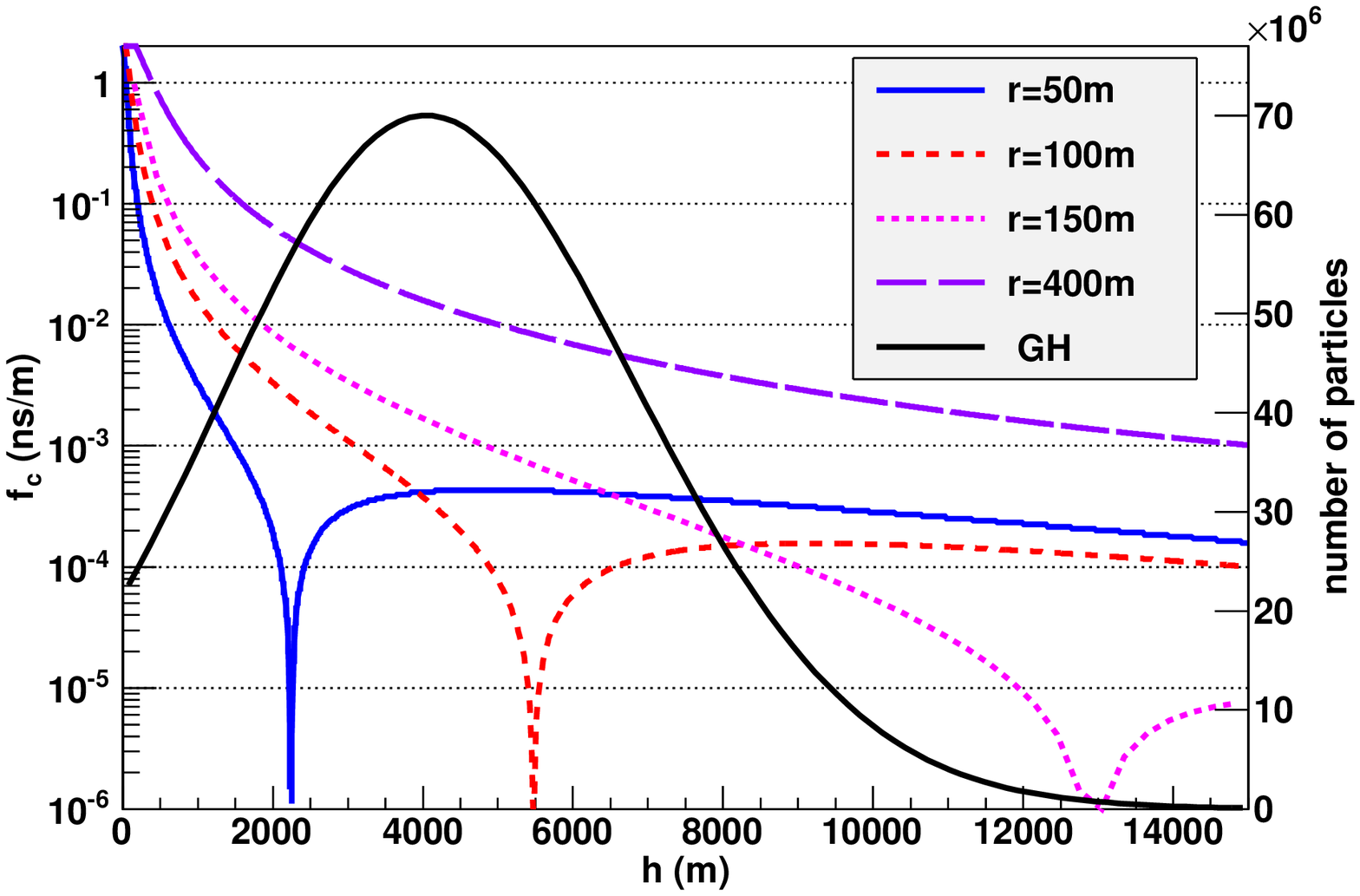}
}
\caption{Left Panel: Compression factor $f_c$ as defined in Eq.~(\ref{eq:compression}) 
for $n=1.0003$ (i.e. the absolute value of the derivative of the curves labeled $n=1.0003$ 
depicted in Fig.~\ref{fig:tdetxdepth}) as a function 
of altitude $h$ for observers at different distances $r$ to the shower axis.
Also shown is the number of charged particles in a $10^{17}$ eV proton shower 
as given by a Gaisser-Hillas (GH) parameterization of the longitudinal shower
development (right axis). Right Panel: Same as left, but for $n=n(h)$.}
\label{fig:fc}
\end{center}
\end{figure}

As stated before, the peak value of the pulse is also determined by other factors besides $f_c$. 
There is a non-trivial interplay between the length and height of the part 
of the shower seen with a small compression factor, the number of particles in that region 
of the shower development, and the distance $R$ to the observer. A clear illustration of this interplay can be seen in Fig.~\ref{fig:fc}, 
where we plot $f_c$ as a function of $h$ along with a Gaisser-Hillas
parameterization of shower development for a $10^{17}$ eV shower. 
When $n=1.0003$ (left panel) the largest compression in time applies to the
region around shower maximum for an observer at $r=100$~m, 
while for an observer at $r=50$~m the largest compression applies to a
portion of the shower with fewer particles, below shower maximum.

One could think that the peak of the pulse will always drop as $r$ increases
and the observer gets further away from shower axis, but the factor
$f_c$ can slow down this trend (e.g. between $r=50$~m and $r=100$~m in
Fig.~\ref{fig:fc}) or even reverse it. Moreover, as $r$ increases, the width of the peak in $f_c$ becomes
larger (this can be appreciated by inspection of the widths at values
of $f_c=10^{-4}$~ns/m in Fig.~\ref{fig:fc}) and this implies that a
larger portion of the shower contributes quasi-simultaneously to the
observed emission. On the other hand, as $r$ increases so does the
distance from the emission point to the antenna, tending to decrease
the signal. 
One can see that even in this simple 1D model the interplay of the
various relevant variables is already very complicated and somewhat
counter-intuitive. For observers at larger distances to the core
(e.g. $r=400$~m in the bottom panel of Fig. \ref{fig:tdetxdepth}), the largest
compression is achieved only at very high altitudes where the number of
particles in the shower is much smaller than at shower maximum,  and thus the
peak value of the pulse is much smaller than that seen by an observer  at
$r=50$~m (bottom panel of Fig.~\ref{fig:tdetxdepth} and Fig.~\ref{fig:fc}).

%
\subsection{Time duration of the electric field pulse}

The 1D model also allows to understand the time duration of the electric
pulse. The shower arrives at ground at $t'_g=0$ and the signal from
the shower when it reaches the ground arrives
at the observer at a time $t_g =  nr/c$. Assuming that the emission is only
due to the shower front, the time
duration $\Delta t$ of the pulse, as seen by the observer, is then given by: 

\begin{equation}
\Delta t=t_g-t_{start} = \frac{r}{c}\left(n-\sqrt{n^2-1}\right) 
\label{eq:delta_t}
\end{equation}
where $t_{start}$ is given by Eq.~(\ref{eq:tdetmin}). This equation is only
valid for
$r$ greater than a few meters\footnote{For very small distances
  $r<h_0(n^2-1)/2n$, then $t_0>t_{g}$ and $\Delta t=t_0-t_{start}$.} and
$r<r_{crit}$ (see Eq.\ref{eq:rcrit}). If $r>r_{crit}$, $t_{start}$
should be replaced by $t_0$, as defined in Eq.~(\ref{eq:t0}), but this does not change
significantly the behavior of $\Delta t$, which stays
approximately linear with $r$. So the observers see an increasingly wider pulse
in time the farther they are from the shower core. Another estimate for the
variation of the pulse width with distance can be found in~\cite{scholtennima2009}.

This approach is an oversimplification as it assumes that the shower still
contributes to the pulse as it reaches ground level, so it needs to be
modified for very inclined showers. In addition it also
neglects the delays of the particles that lag behind the shower front and
also contribute to the pulse, making it wider. Still
the simple expressions obtained can be quite useful to interpret the
results from the full simulation, shown in the next section.

\subsection{Dependence on refractive index}
\label{sec:1Dmodel-n}
The 1D model also allows to study the dependence of the pulse properties on the
index of refraction. For the purpose of understanding the relevance of $n$ on
the pulse, we first assume $n$ to be constant with height. 
Using Eq.~(\ref{eq:tdetmin}) it is straightforward to deduce that the start-time
of the pulse increases with $n$ because the propagation time of the signal from 
the height of emission to the observer increases with $n$. Furthermore, the
time duration of the pulse also depends on $n$. As can be seen in
Eq.~(\ref{eq:delta_t}), the time duration of the pulse decreases slightly with
increasing $n$.

\begin{figure}
\begin{center}
\scalebox{0.65}{
\includegraphics{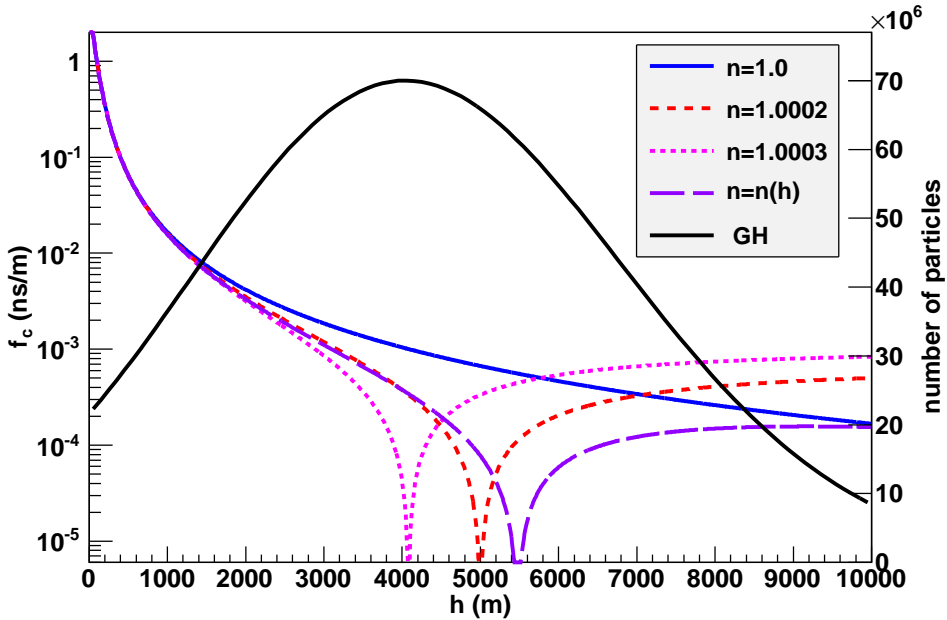}
}
\caption{Same as Fig.~\ref{fig:fc}, but for an observer at $r=100$~m and
  several values of $n$, including a model for the variation of $n$ with
  altitude $n(h)$ (see Section~\ref{sec:n(h)}).}
\label{fig:fc-varn}
\end{center}
\end{figure}

Another dependence of the pulse properties on $n$ arises because the factor $f_c$ is also
dependent on $n$. An observer at a fixed given distance $r$ from the core
will see different parts of the same shower with a small $f_c$ (i.e. with a large
compression in time) for different values of $n$, since $h_{start}$
(which defines the altitude at which $f_c=0$ and the shower is seen
with $\theta=\theta_C$) decreases with $n$. To illustrate this we show
in Fig.~\ref{fig:fc-varn} the factor $f_c$ as a function of $h$ for an
observer at $r=100$~m for several values of $n$. One can see that as
$n$ decreases the height at which $f_c=0$ (i.e. $h_{start}$) increases,
leading to different parts of the shower being largely compressed in time. The
effect on pulse height will depend on the number of particles at
$h_{start}$, which in turn depends on the distance $r$ to the
observer. For $r<100$~m (not shown in Fig.~\ref{fig:fc-varn}),
the highly compressed part of the shower for $n=1.0003$ is below
shower maximum. As $n$ decreases, the compressed part of the shower moves towards the maximum, tending to increase the
pulse. On the other hand, for $r>100$~m, the height $h_{start}$ is
above the maximum, and a decrease in
$n$ will move the compressed part of the shower further away from the maximum,
tending to decrease the pulse height. For large values of $r$,
$h_{start}$ is in a region with very few particles, decreasing  the effect of
a low $f_c$ in pulse height. Furthermore, in this
part of the shower the number of particles increases only very slowly with
decreasing $h$, since the density at these altitudes is low, and the effects
of changes in $n$ on pulse height become less important.

As discussed before, relativistic effects such as seeing the later
parts of the shower before the earlier ones and seeing two parts of the shower
simultaneously can only be observed at distances $r<r_{crit}$ from the
core. Since $r_{crit}$ increases with $n$, observers further from the core
will also see these effects as $n$ increases.

\section{ZHAireS simulations}
\label{sec:results}
In this section we present the electric field as obtained in ZHAireS
simulations of atmospheric showers. In particular
in this section we will obtain the behavior of different features of the time
pulse with $r$ and refractive index, but this time in realistic simulations 
that take into account the full complexity of atmospheric showers. We will
see that the behavior follows the qualitative behavior obtained with
the 1D model. At the end of this section we also show some results of the
electric field calculated in the frequency domain and compare them with
Fourier transforms of the time pulses.  

\subsection{Dependence of the electric field pulse on distance to shower axis}

\begin{figure}
\begin{center}
\scalebox{0.65}{
\includegraphics{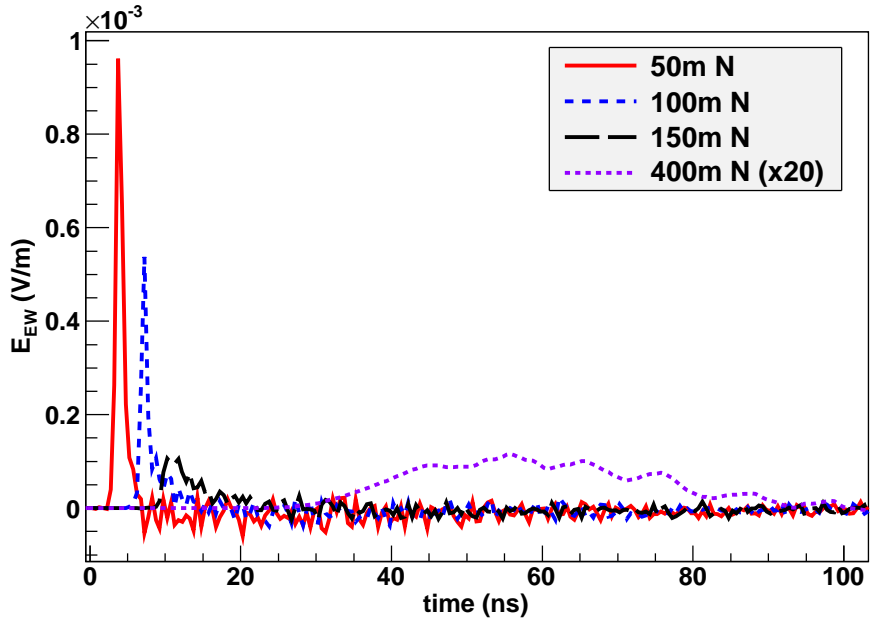}}
\caption{EW component of the electric field at $r=50$, 100, 150 and 400 m
northwards of the shower core as obtained in ZHAireS simulations of a $10^{17}$ eV 
proton-induced vertical atmospheric shower. These simulations were performed 
with the exponential model for the variation with altitude of the refractive index.
Note that the signal at $r=400$~m is arbitrarily multiplied by a factor 20 for plotting purposes.}
\label{fig:field_dep_r}
\end{center}
\end{figure}

In Fig.~\ref{fig:field_dep_r} we show the East-West (EW) electric field
component as a function of time obtained from simulations of $10^{17}$ eV vertical atmospheric showers induced
by protons. We used a horizontal magnetic field $|\vec{B}|=23\mu T$ pointing
north for this particular simulation. The electric field is shown at different distances $r$ northwards from the 
shower core. As predicted by the 1D model in Eqs.~(\ref{eq:tdetmin})
and (\ref{eq:delta_t}), it can be seen that 
the start-time of the field and the duration in time (at least the easily
visible positive part of the pulse) both increase 
with $r$ for the observers shown in the plot. 

\begin{figure}
\begin{center}
\scalebox{0.65}{
\includegraphics{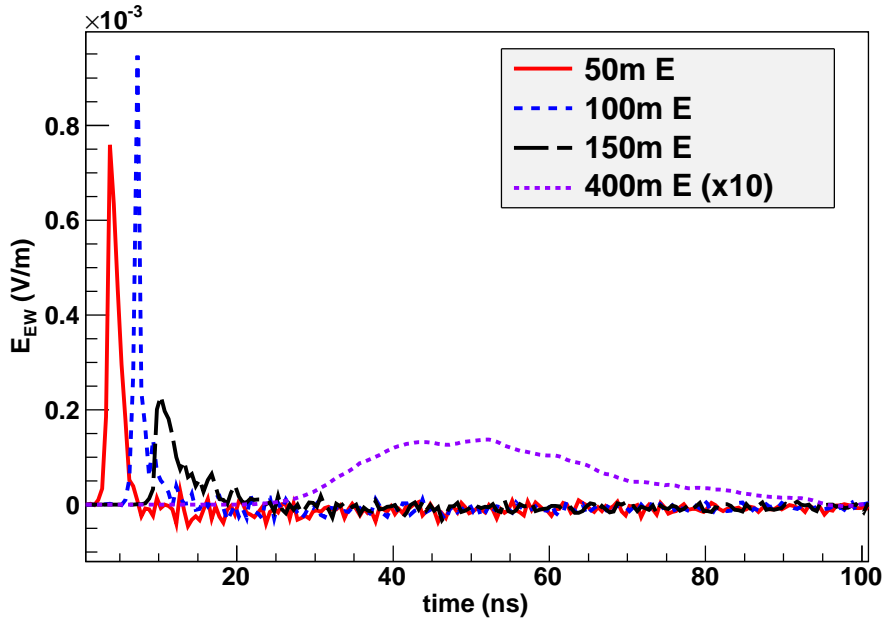}}
\caption{Same as Fig.~\ref{fig:field_dep_r}, but for observers east of the core.}
\label{fig:field_dep_r-E}
\end{center}
\end{figure}

The peak of the electric field decreases with $r$ for an observer North of the
shower core, but not in a linear manner. In fact it decreases roughly as $r$
from $50~~$m to $100$~m, but faster than $r$ from $100$~m to $150$~m, as can be
seen in Fig.~\ref{fig:field_dep_r}. As explained above, there is a non-trivial interplay between the distance from the shower to the
observer, the factor $f_c$ and the number of particles in
the region of the shower seen with a large compression in time, illustrated 
in the right panel of Fig.~\ref{fig:fc}. The number of particles at $h_{start}$
increases only slightly from $r=50$~m to $r=100$~m, but decreases by an order of
magnitude between $50$~m and $150$~m. Since the distance $R$ from emission
point to observer increases roughly as $R\sim r$, it is twice (three times) as
big at $100$~m ($150$~m) than at $50$~m. This explains why the height of the peak at
$100$~m is about half the height at $50$~m, while it is much smaller at $150$~m. 
We have also observed in our simulations that the relative height of the
pulse at different distances also depends on the direction of the observer
w.r.t. the shower core. In Fig.~\ref{fig:field_dep_r-E} we show the emission
from a similar shower as in Fig.~\ref{fig:field_dep_r}, but for observers East
of the core. For an observer East of the shower core, the peak amplitude appears to
be maximal at around $r=100$~m, a result in principle compatible with
\cite{scholtenprl}. Also, the decrease in pulse height with $r$ seems to be
slower in Fig.~\ref{fig:field_dep_r-E}, when compared to the observer North of
the core (Fig.~\ref{fig:field_dep_r}). A similar dependence on antenna
position can also be seen for the spectra at 1 MHz, shown in Fig.\ref{fig:ldfplpot}. This difference is, in part, due to the interference of the
geomagnetic and Askaryan components of the emission, as will be discussed in
section \ref{sec:polarization}.

\subsection{Dependence of the electric field pulse on the refractive index}

\begin{figure}
\begin{center}
\scalebox{0.35}{
\includegraphics{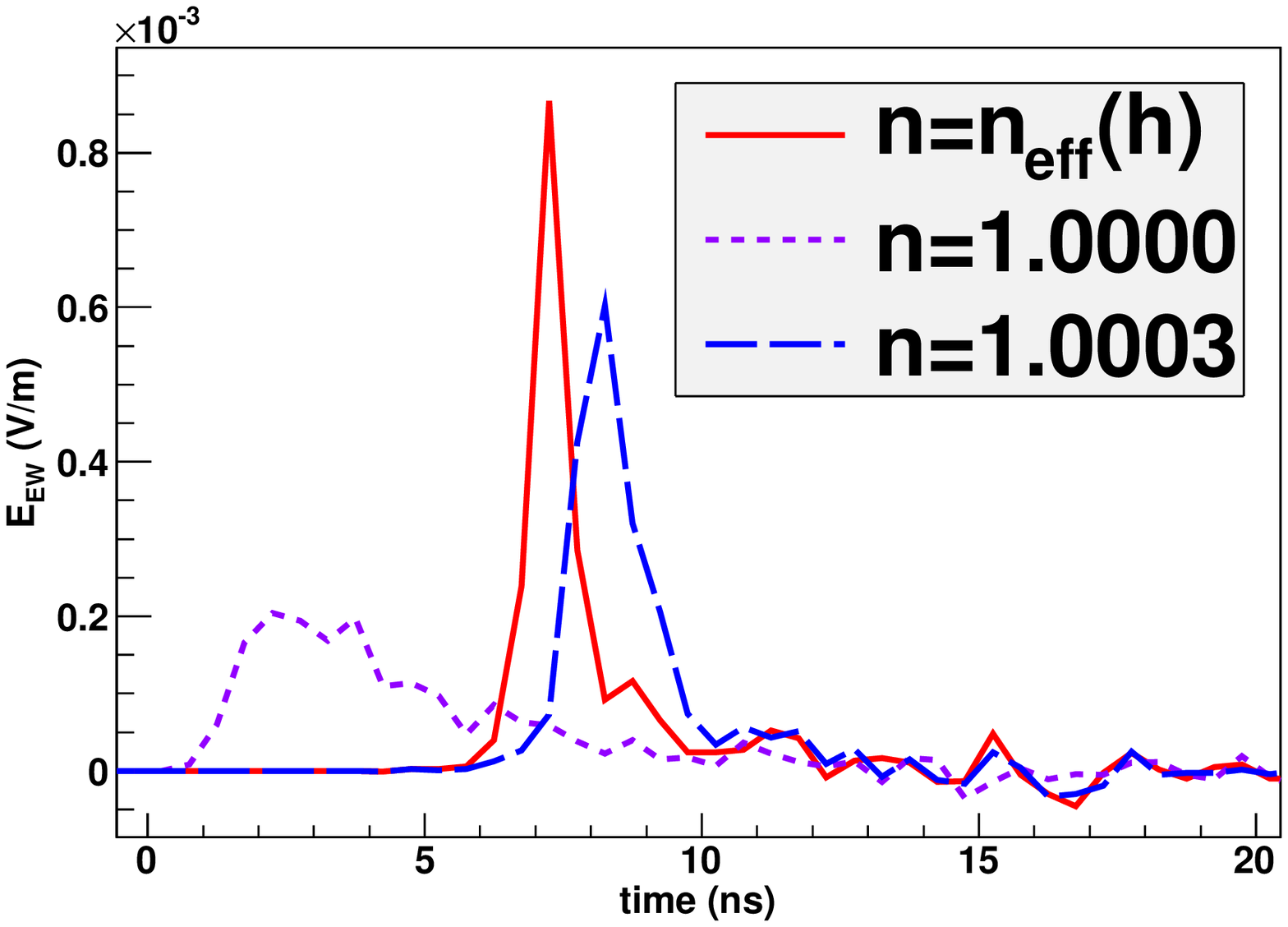} \includegraphics{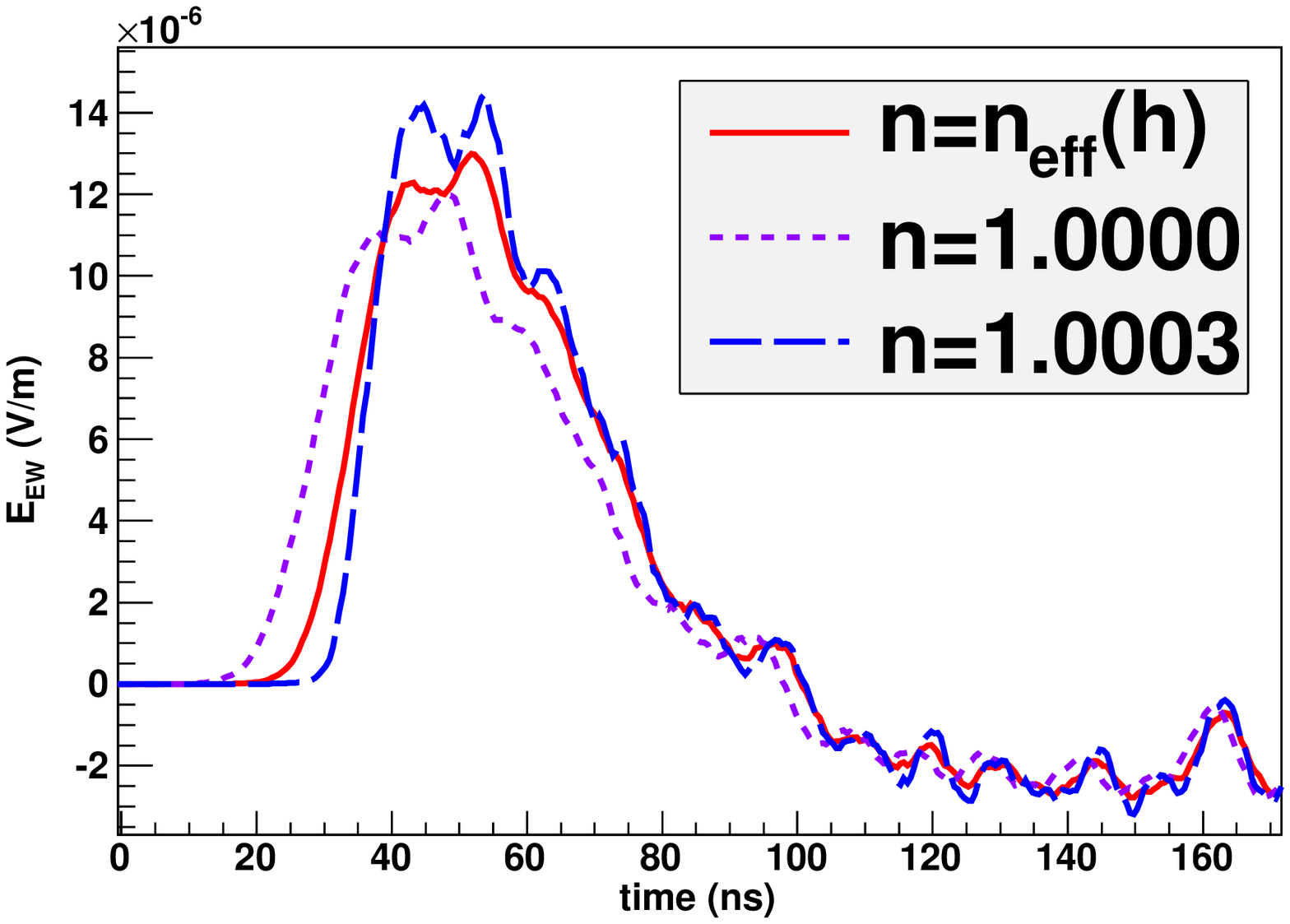}}
\caption{EW component of the electric field at $r=100$~m (left panel) and $r=400$~m (right panel)
eastwards from the core for a vertical $10^{17}$ eV proton-induced shower as obtained in ZHAireS 
simulations. The calculation was done using several refractive indexes.}
\label{fig:ncompallvert}
\end{center}
\end{figure}

In Fig.~\ref{fig:ncompallvert} we show the EW component of the electric field as a function of time obtained from simulations of $10^{17}$ eV vertical atmospheric showers induced
by protons, using a magnetic field $|\vec{B}|=23\;\mu T$ with an inclination of
$-37^\circ$ and a declination of $0^\circ$. The calculations were performed with two constant refractive
indices, namely $n=1.0$ and $n=1.0003$, as well as with a refractive index
varying with altitude according to Eq.~(\ref{eq:n(h)}).

Firstly, the larger the refractive index the later the pulse starts as
obtained with the 1D model above (see Eq.~\ref{eq:tdetmin}). The larger the $n$ the 
smaller the propagation speed, and so the signal reaches the observer at later times. 
For the refractive index changing with altitude, which varies between $n=1.0$
and $n=1.0003$, the start-time falls between the two obtained for the two
constant refractive indices, as expected. 

For observers close to the shower axis ($r=100$~m in the left panel of
Fig.~\ref{fig:ncompallvert}), small changes in $n$ are responsible for large changes in
peak height and width. This can be qualitatively understood, as described in
Sections~\ref{sec:1Dmodel-peak} and \ref{sec:1Dmodel-n}, in terms of the
non-trivial interplay between the various geometrical factors. In Fig.~\ref{fig:fc}
and \ref{fig:fc-varn} one can see that for $r=100$~m the number of particles
in the low $f_c$ region changes only slightly from $\sim 70\cdot 10^6$ when $n=1.0003$
to $\sim 60\cdot 10^6$ when $n=n(h)$, while the length of the shower seen with
$f_c<10^{-4}$ doubles. For observers further away from the core ($r=400$~m in
the right panel of Fig.~\ref{fig:ncompallvert}), the effect of the varying $n$
in pulse height is much less pronounced, since $f_c\sim 0$ only above the
shower, as can be seen in the bottom panel of Fig.~\ref{fig:tdetxdepth}. This
will make the compression factor in the shower region very similar for
different values of $n$, as illustrated by the fact that the curve labeled
$r=400$~m in Fig.~\ref{fig:fc} changes only very slightly  from the left panel
($n=1.0003$) to the right panel ($n=n(h)$).
%

\subsection{Dependence of the electric field pulse on the shower zenithal angle} 
\label{sec:vertinclined}

The features of the electric field at a fixed distance to the shower
axis are expected to depend strongly on the shower zenithal angle. 
The main reason for this is that the shape of the curve relating 
the observer time and the source depth - which largely determines
the characteristics of the pulses as explained above - depends 
strongly on the geometry of the system formed by the observer and the shower.
The electric field from a certain region of the shower depends on 
the angle $\theta_i$ (with which the observer sees that region) through 
the factor $\vert 1-n\beta\cos\theta_i\vert$ in 
Eq.~(\ref{eq:timefresnel}), on the number of particles $N_i$ in the
region, and on its distance to the observer, $R_i$. For $\theta_i$ close 
to the Cherenkov angle, the emission is enlarged with respect to other angles,
since $\vert 1-n\cos\theta_i\vert$ is small. 

\begin{figure}
\begin{center}
\scalebox{0.65}{
\includegraphics{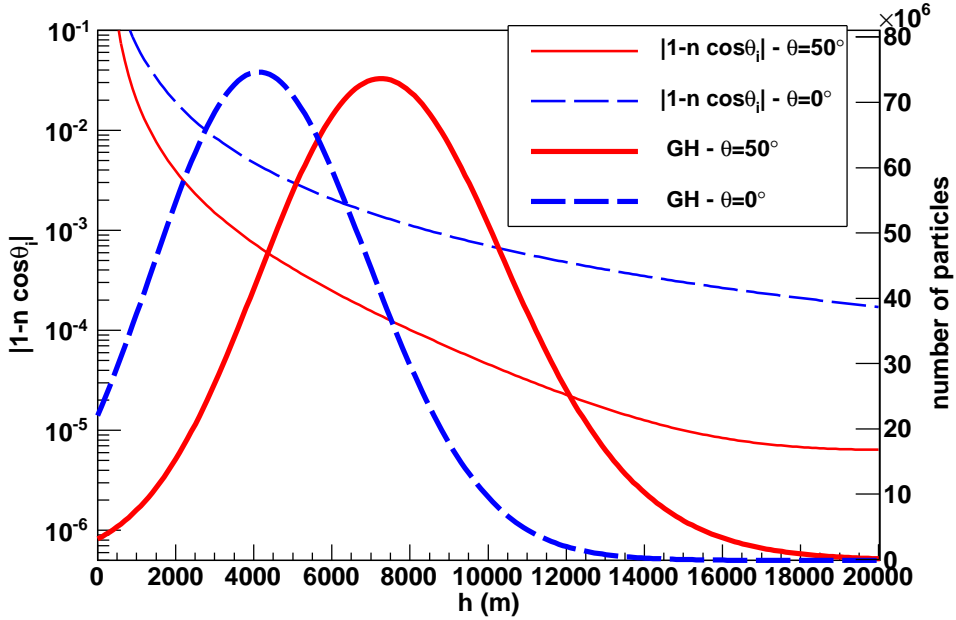}}
\caption{Geometrical angular term $|1-n\cos{\theta_i}|$ as a function of altitude $h$
  for a vertical shower (dashed blue line) and a shower with zenithal
  angle $\theta=50^{\circ}$ (solid red line). Also shown are Gaisser-Hillas parameterizations of the number of particles (linear right axis) as a
  function of height for a 100PeV vertical shower (thick
  dashed blue line) and an equivalent one with $\theta=50^{\circ}$ (thick solid red line). The
  calculations were done numerically for $n=n(h)$ and $r=400$~m (early
  part of inclined shower), using an 1D model that allows non-vertical
  showers, similar to the one described for vertical showers. }
\label{fig:vert_vs_inclined}
\end{center}
\end{figure}

The interplay between  $\vert 1-n\cos\theta_i\vert$ (or equivalently
the factor $f_c$) and $N_i$ is shown 
in Fig.~\ref{fig:vert_vs_inclined} for a vertical (zenithal angle $\theta=0^\circ$)
and an inclined shower ($\theta=50^\circ$). The curves in Fig.~\ref{fig:vert_vs_inclined} 
were obtained for an observer at a distance $r=400$~m (in the early part of the inclined
shower) and $n=n(h)$ with a modified 1D model, similar to the one developed in
Section \ref{sec:1Dmodel}, but which can handle inclined
showers. Fig.~\ref{fig:vert_vs_inclined} suggests that inclined showers
produce larger signals than vertical showers for distances greater than a
couple of hundred meters from the core. The main reason for this is that at larger distances to the shower axis, inclined showers are viewed with angles closer to the Cherenkov angle than vertical showers,
and so the angular factor $\vert 1-n\beta\cos\theta_i\vert$ is smaller,
boosting the emission and the net signal from inclined showers. A similar conclusion was also reached in~\cite{Gousset_inclined}.

\begin{figure}
\begin{center}
\scalebox{0.35}{
\includegraphics{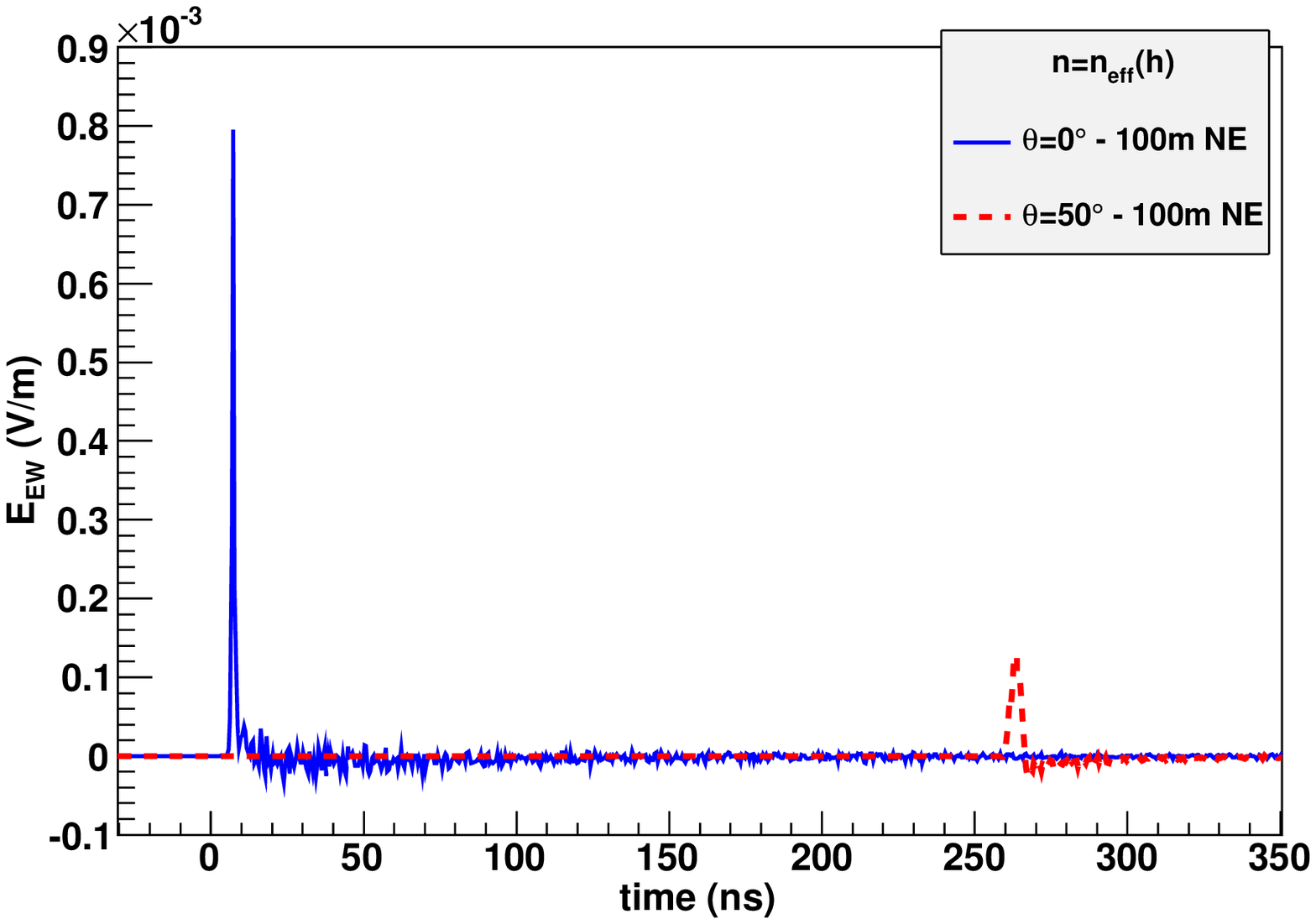}\includegraphics{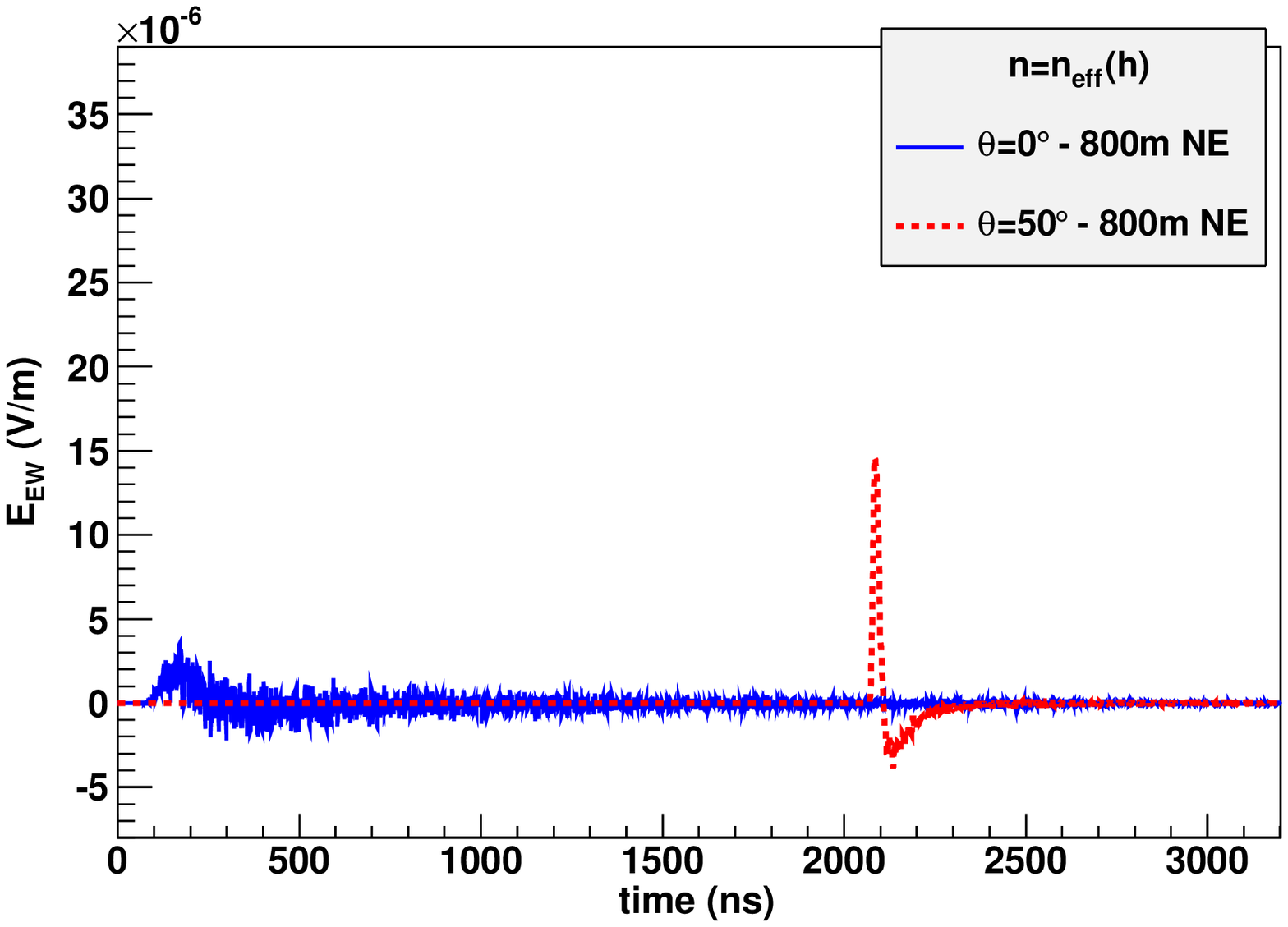}}
\caption{Comparison between the electric field as a function of time as obtained in 
ZHAireS simulations of $10^{17}$ eV proton-induced showers with zenithal angle 
  $\theta=0^\circ$ (blue solid) and $\theta=50^\circ$ (red dashed), for antennas at 
  $r=100$~m (left) and $r=800$~m (right) NE of the shower core on the ground. 
  The inclined shower comes from the SE. The simulations were done using a variable
  refractive index model for the atmosphere.}
\label{fig:compvert-inclined-nvar}
\end{center}
\end{figure}

Using ZHAireS, we investigated the dependence of the field on zenithal angle
as a function of distance to shower axis, in a full Monte Carlo simulation
of the shower development in air with a realistic model for the refractive index. 
We compared the signals obtained from $10^{17}$ eV proton-induced showers with
$\theta=0^\circ$ and $\theta=50^\circ$. We found that the results obtained with
ZHAireS are in qualitative agreement with our 1D model and with the calculations 
in~\cite{Gousset_inclined}. In Fig.~\ref{fig:compvert-inclined-nvar} we compare 
the East-West components of the electric field for the vertical and inclined showers 
for an antenna at $r=100$~m (left) and $r=800$~m (right) located in the North-East 
of the shower core. Clearly at large distances to the core the peak value of the
electric field in the inclined shower is much larger than that of the vertical 
shower, a factor $\sim 5$ in this particular simulation. This trend of an increased non-vertical shower signal at large distances from the core 
is present in all directions with respect to the shower core. 
Note also the shift in the start-time of the pulse in the inclined shower
w.r.t. the vertical one, due to the different geometry of both showers.

\begin{figure}
\begin{center}
\scalebox{0.35}{
\includegraphics{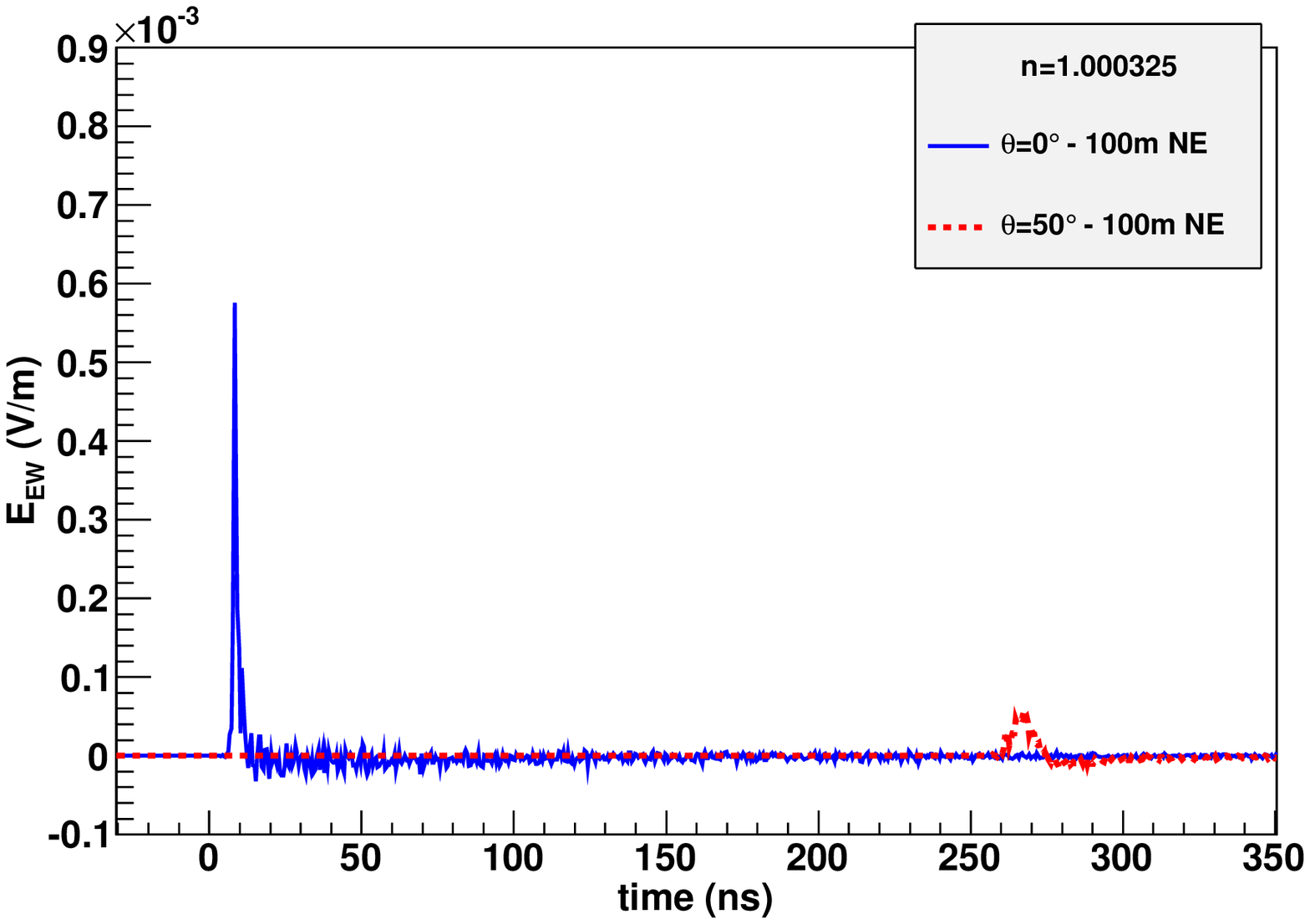}\includegraphics{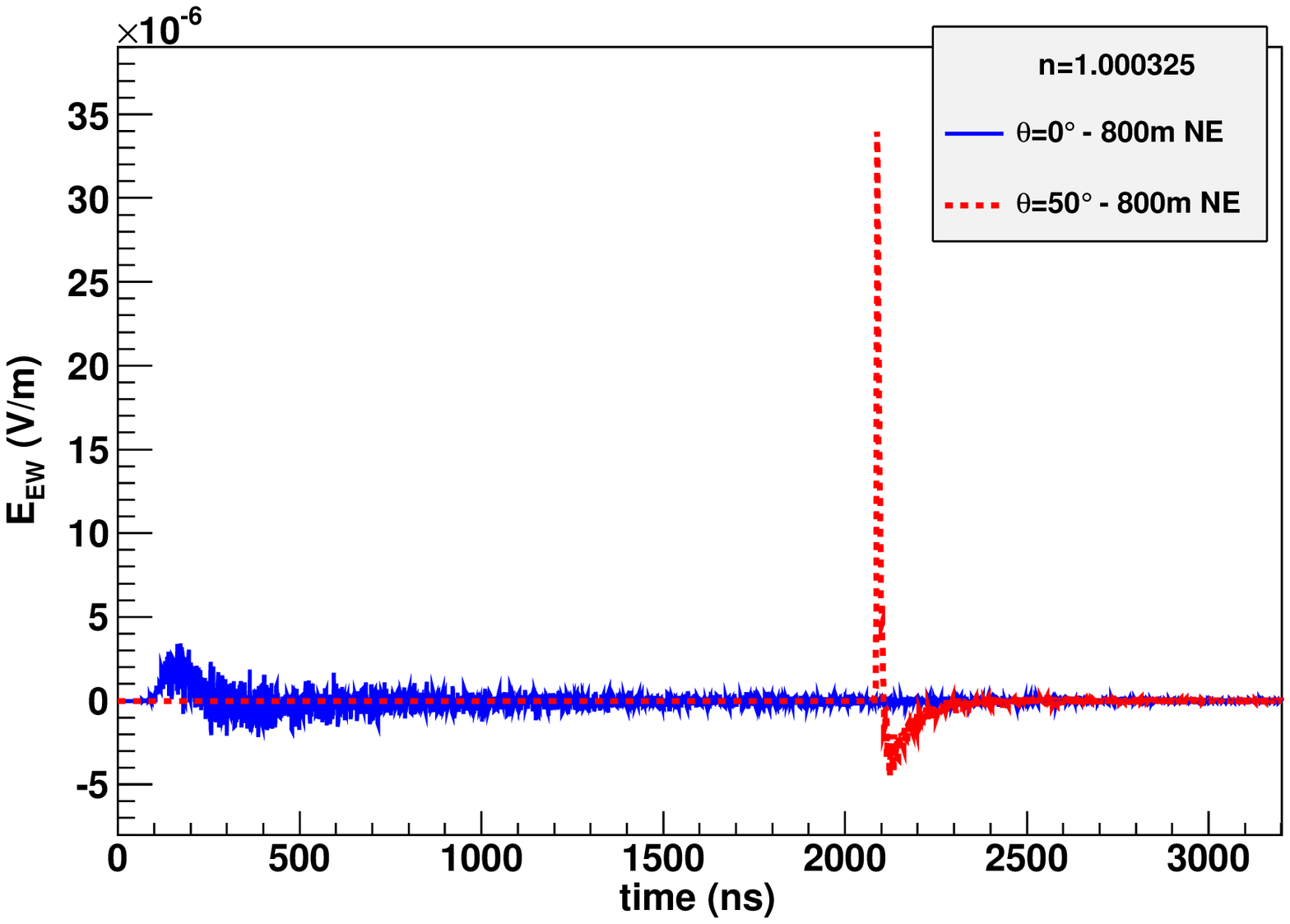}}
\caption{Same as Fig. \ref{fig:compvert-inclined-nvar}, but using a constant 
  refractive index $n=1.00035$ for the whole atmosphere.}
\label{fig:compvert-inclined-nair}
\end{center}
\end{figure}

We have also found that the signal in non-vertical showers at large
distances from the core is very sensitive to the decrease of the
refractive index with altitude. In Fig. \ref{fig:compvert-inclined-nair} 
we plot the electric field for the same showers as in 
Fig. \ref{fig:compvert-inclined-nvar}, but instead of the model of variable refractive
index with altitude, the fields shown were calculated using a constant refractive index 
$n=1.000325$. It can be clearly seen in the left panel of
Fig. \ref{fig:compvert-inclined-nair} that for the antenna at
$r=100$~m North-East from the core,
the use of a constant $n$ decreased both the vertical and non-vertical EW field
component w.r.t. the variable $n$, while at $r=800$~m (right panel), the
signal in the vertical shower stayed practically unchanged and the peak in 
the non-vertical shower more than doubled in value. Again this behavior
can be traced back to the geometrical dependence of the interplay between 
the various key elements in the calculation of the electric field.  
%

\subsection{Radio pulses in the frequency domain}
\label{sec:resultsfreq}

\begin{figure}
\begin{center}
\scalebox{0.65}{
\includegraphics{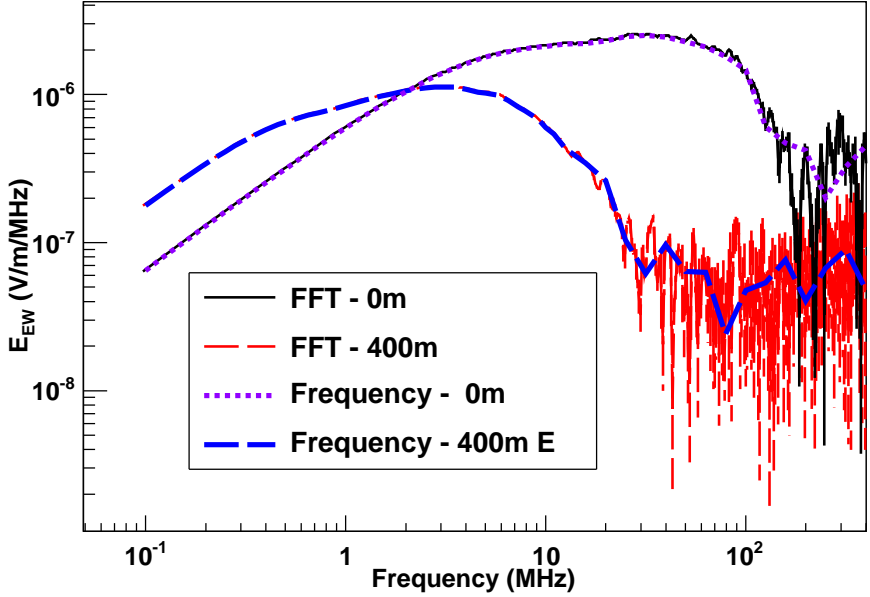} }
\caption{Comparison between the emission spectrum calculated by ZHAireS in the
frequency domain, and the FFT of the time domain signal for the same 100 PeV vertical shower.}
\label{fig:fft}
\end{center}
\end{figure}

 In Fig. \ref{fig:fft} we compare the results of ZHAireS frequency domain
calculations with Fast Fourier Transforms (FFT) of the calculated time
domain signals for the same shower. The convention used for the normalization
of the Fourier transform is in Eq.~(\ref{eq:ft}). One can see that the
agreement between the spectra and the FFT is very good up to high
frequencies, at which the width of the positive peak in the time domain becomes
important, since it is the smallest large scale structure of the time pulse. More quantitatively, above $\nu\sim 1/ \Delta T$, where $\Delta T$
is the characteristic width of the peak of the pulse in time, the FFT starts fluctuating. For $0$~m ($400$~m), the peak width is around $5$~ns ($50$~ns), 
leading to large fluctuations starting at around $200$~MHz
($20$~MHz)\footnote{Similar fluctuations can also be seen
  in~\cite{REAS3}.}. This high frequency part of the spectrum is largely incoherent
and sensitive to shower to shower fluctuations and to the thinning level used
in the simulation. These
fluctuations are partly physical (incoherence level, as will be discussed
below) and partly unphysical (bin size effects, thinning, FFT numerical error,
etc). Also, in the direct frequency domain calculations they are slightly smaller than
in the Fourier-transform of the time pulse\footnote{Note that due to computing time issues, the frequency bin
  used in the frequency domain simulations shown in Fig.~\ref{fig:fft} is much
  wider than the frequency bin used in the FFT, making the fluctuations appear much smaller
  than they really are.}. This means that to study the spectrum at high
frequencies it is best to calculate shower emission directly in the frequency
domain, reducing the unphysical contributions to the fluctuations.

The spectrum of the radio emission depends on the distance $r$ from the core on
the ground to the antenna. As can be seen in figure \ref{fig:fft}, the
frequency at which the Fourier components reach a maximum decreases from $\sim 30$~MHz at $r=0$~m to $\sim 3$~MHz at
$400$~m. This cutoff
frequency is related to the time duration
of the pulse for a specific observer and serves as a boundary between the fully coherent and incoherent parts of the
spectrum. Above this frequency, the emission is subject to shower to
shower fluctuations. At even higher frequencies, effects due to thinning in the
simulation become important, since the spectrum becomes sensitive to the fine
details of the shower, even down to single
particles for the highest frequencies (not shown). A 2D model that explains this behavior in the Fraunhofer regime,
but which is still relevant, can
be found in~\cite{alvz06}. 

\begin{figure}
\begin{center}
\scalebox{0.35}{
\includegraphics{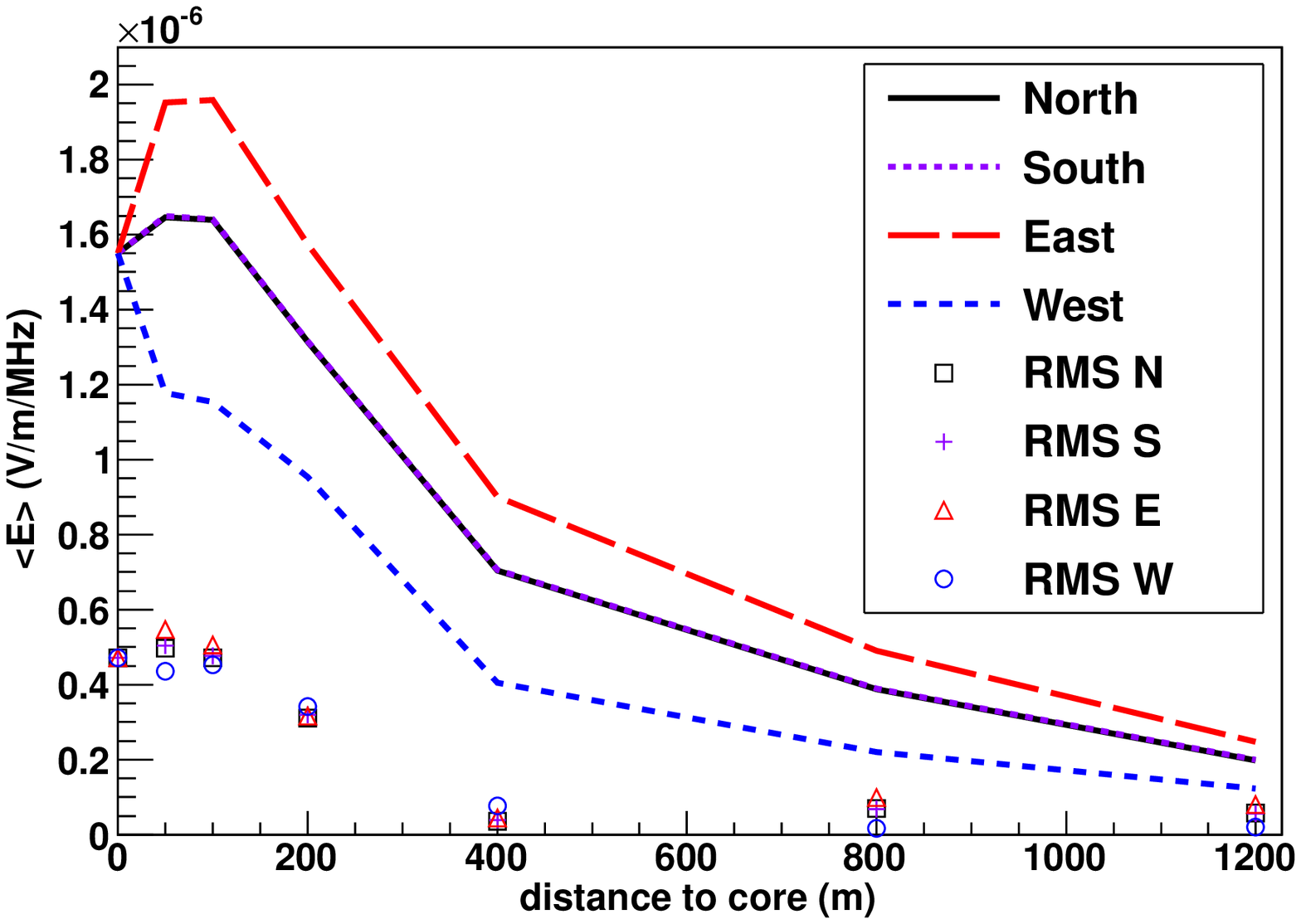} \includegraphics{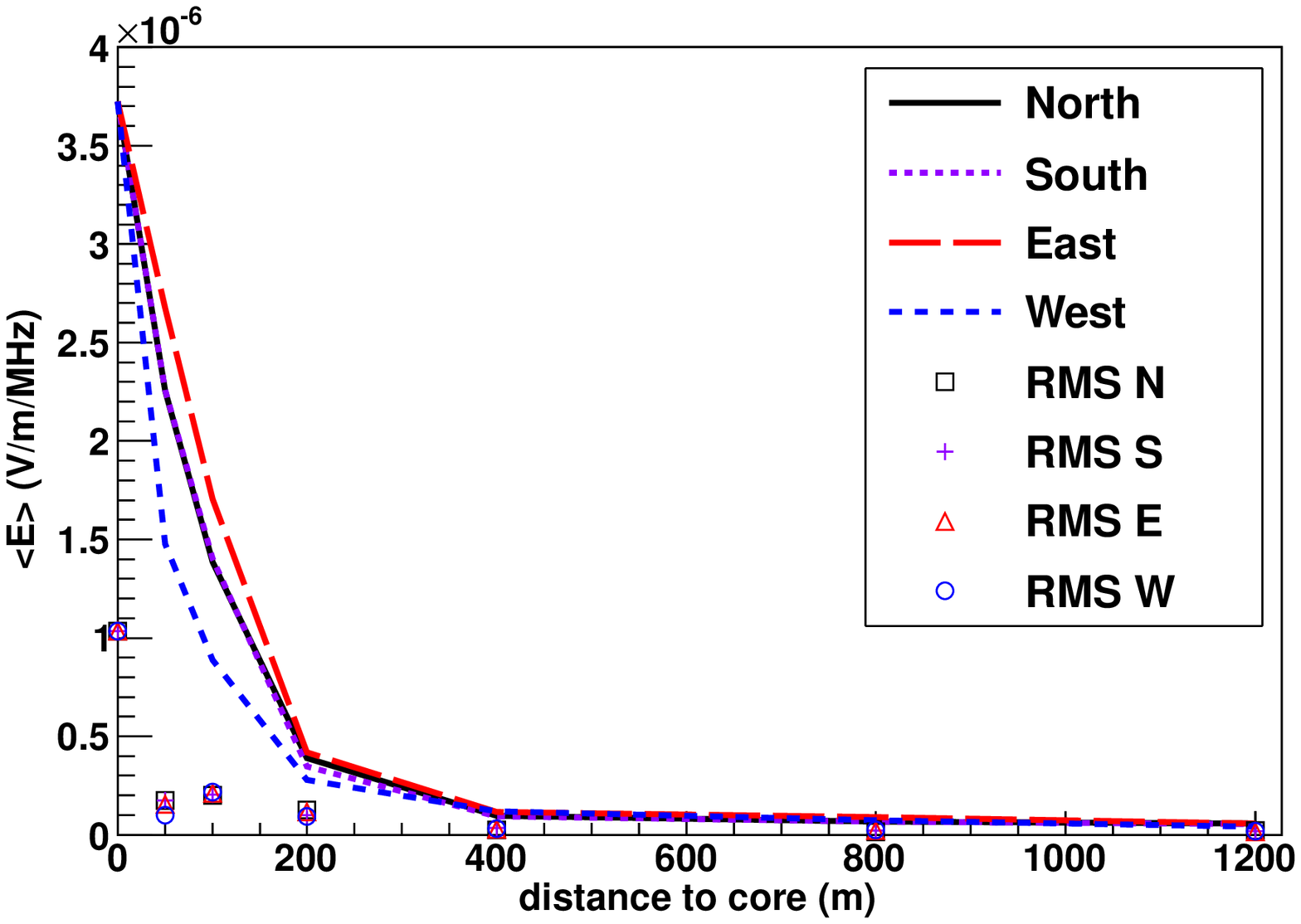}}
\caption{Average electric field vs distance to core for 10 vertical showers of
  100 PeV simulated with ZHAireS at $1$~MHz (left) and $60$~MHz (right). The
  observers are placed at positions N, S, E and W of the shower core. Note that for
  observers N and S the fields are similar and lie on top of each other in the
  figure. Also shown as points are the RMS of the electric field.}
\label{fig:ldfplpot}
\end{center}
\end{figure}

The fact that the cutoff frequency of the emission decreases with
distance to the core has implications on the measured lateral distribution of the
signal. Depending on the measured frequency range, the Fourier components of
the signal will drop
at different rates as the distance increases. In Fig. \ref{fig:ldfplpot} we show
the average total field as a function of distance to the core at $1$~MHz
(left) and at $60$~MHz
(right), along with their RMS, calculated from simulations of 10 proton
showers of $100$ PeV. At $60$~MHz, i.e. in the incoherent part of the
spectrum, the field decreases much faster with
increasing distance to the core compared to the coherent part of the
spectrum at $1$~MHz. This behavior can be relevant for the design of radio
detector arrays. An explanation of the reason why the largest
(smallest) pulses are received at antennas located in the East (West)
of the shower is given in the next section.

At the fully coherent 1 MHz frequency (Fig. \ref{fig:ldfplpot} left), one can also see that a
maximum appears at around $r\sim100$ m. A similar maximum was reported in
\cite{scholtenprl}, but for full bandwidth pulse height. Furthermore, Fig. \ref{fig:ldfplpot} suggests that this
maximum is dependant not only on frequency, but also on observer direction
w.r.t. the shower core, disappearing for antennas to the west. Our full
bandwidth pulses (Figs. \ref{fig:field_dep_r} and \ref{fig:field_dep_r-E}) show
a similar dependence of the maximum of the emission with observer direction. This suggests that the mechanism responsible for this maximum is weaker
than the interference effect between the geomagnetic and Askaryan components
of the emission, which is discussed in the next section. 
%

\section{Polarization properties of the signal and asymmetries}
\label{sec:polarization}

\begin{figure}
\begin{center}
\scalebox{0.55}{
\includegraphics{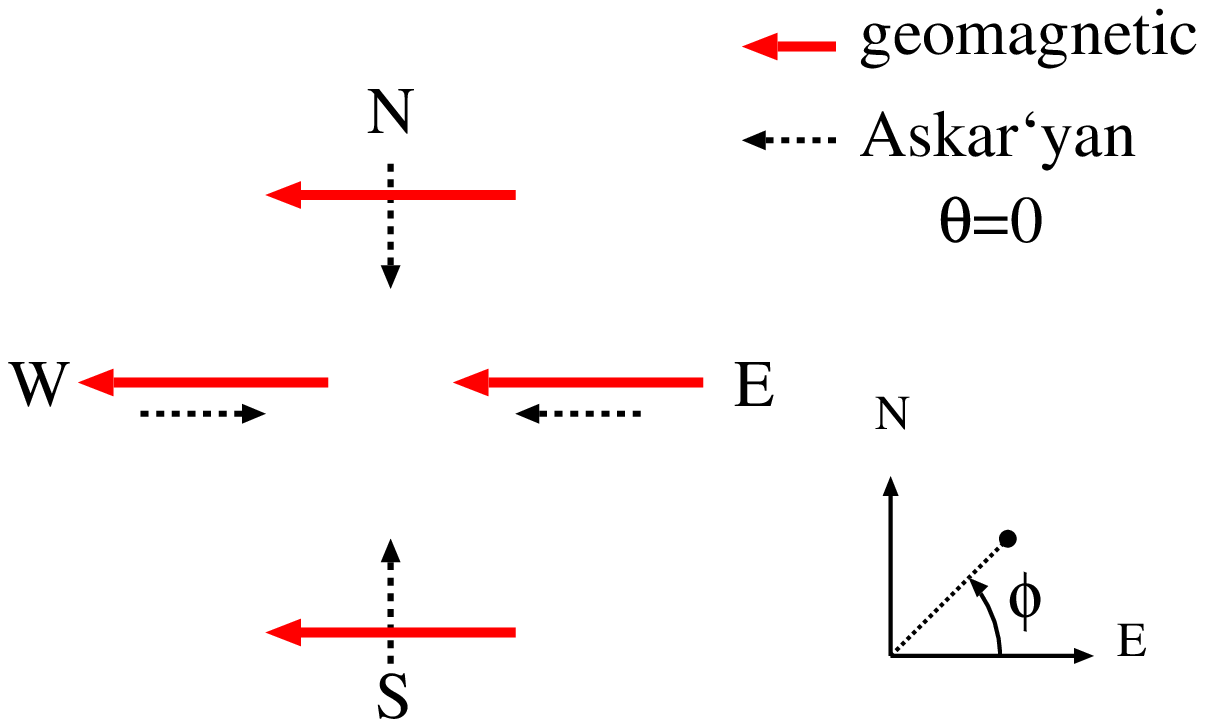}}
\scalebox{0.35}{
\includegraphics{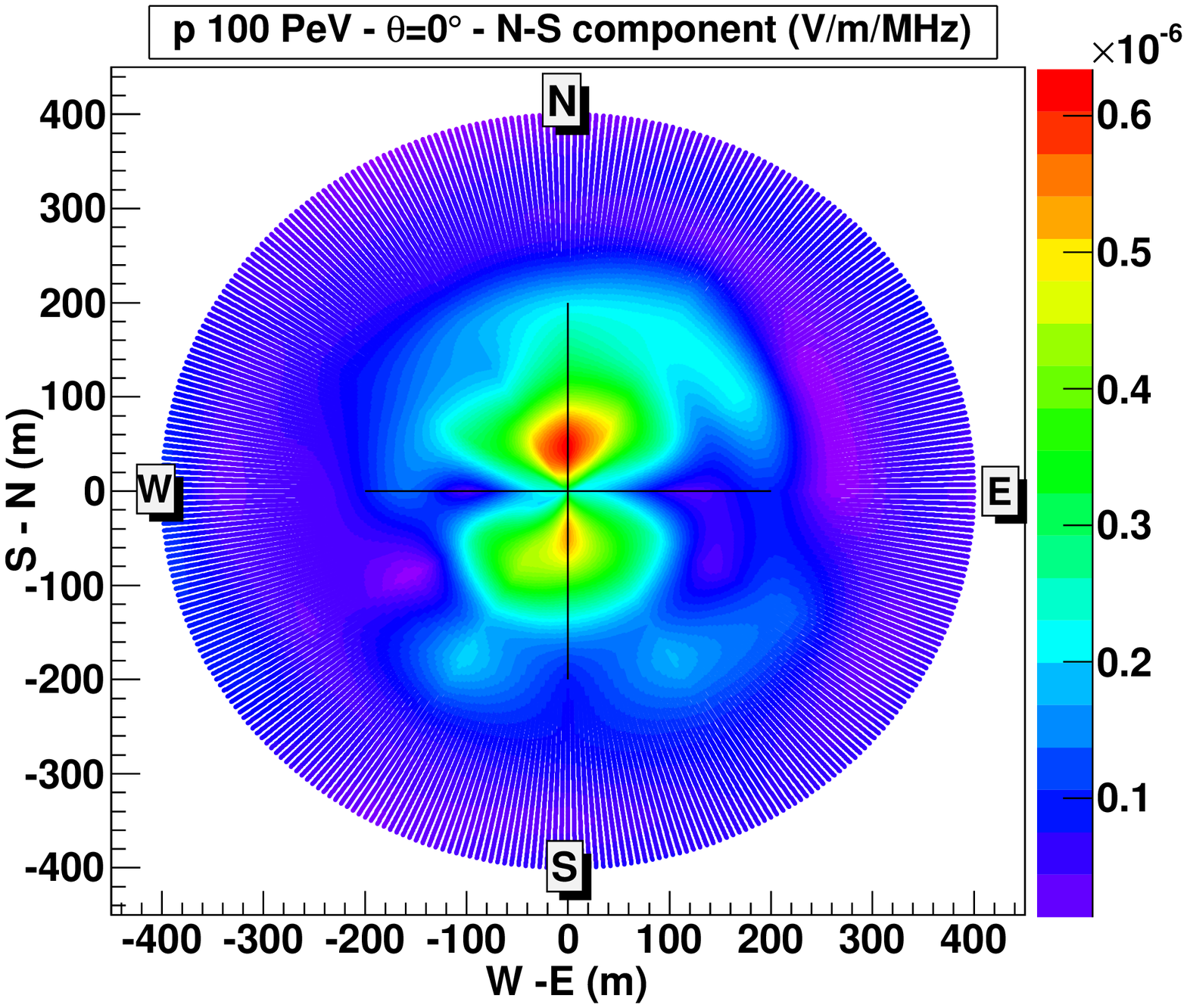}\includegraphics{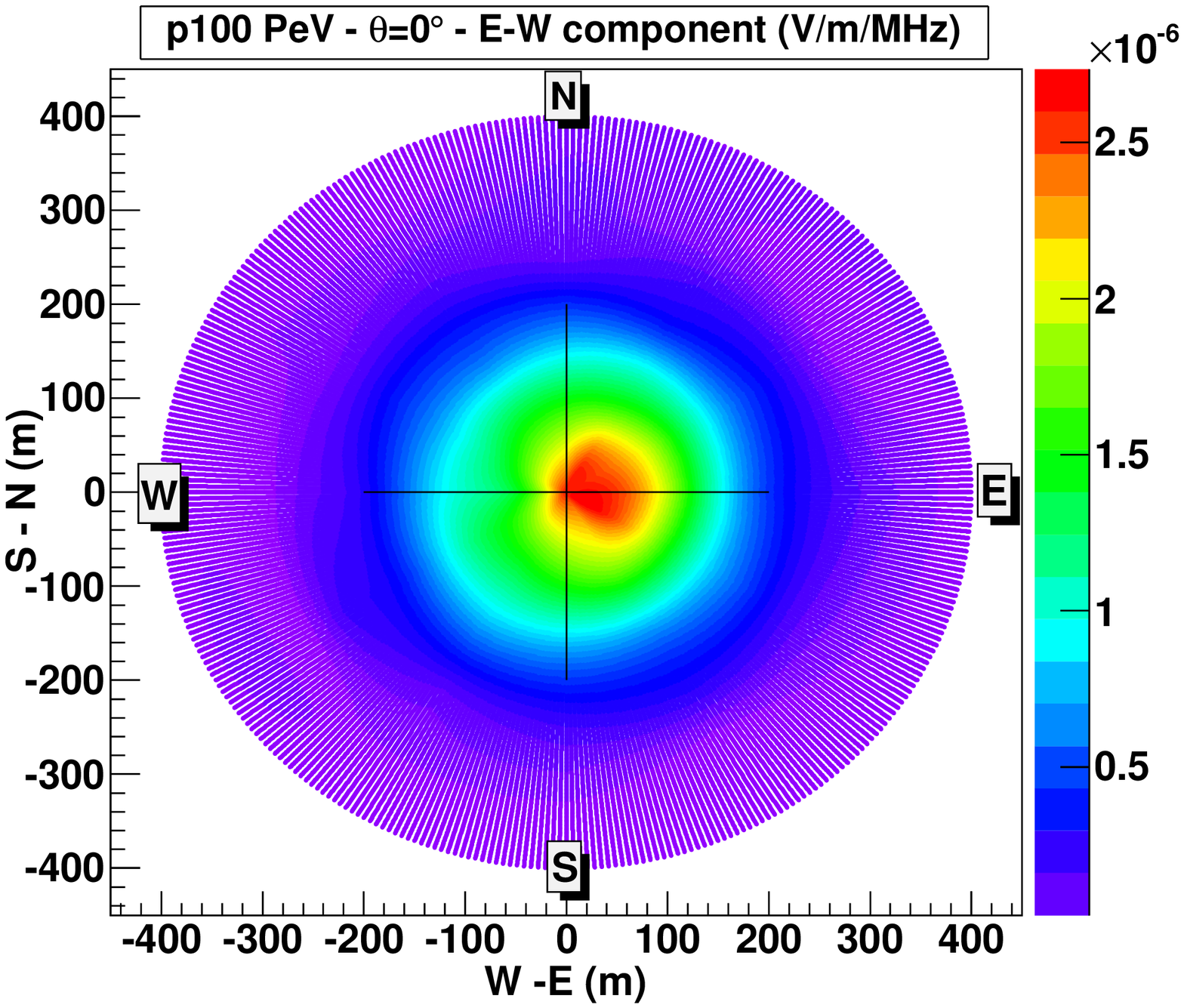}}
\caption{(Color online) Top: Sketch of the projection on the ground of the electric field
induced by the Askaryan and geomagnetic emission mechanisms in 
a vertical shower $\theta=0^\circ$ (see also \cite{REAS3}).
Bottom: Left panel: North-South component of the electric field
$E_{NS}$ as a function of the position around the shower core 
as obtained in ZHAireS simulations of vertical showers induced
by protons of energy $10^{17} eV$. Right panel: East-West component
$E_{EW}$ of the electric field obtained in the same simulations.
The color scale indicates the magnitude of the components of the
field in ($V/m/MHz$), note the different scale in the left and right
panels.}
\label{fig:polvertical}
\end{center}
\end{figure}

The polarization of the electric field depends on the
relative importance of the two main mechanisms thought to be responsible 
for the radio emission~\cite{Scholten_MGMR_sims}. Cherenkov radio emission due to
the excess of electrons in the shower (Askaryan effect) has a characteristic
axial polarization in relation to the shower axis. On the other hand, the
polarization of the emission due to geomagnetic follows
$-\vec{v}\times\vec{B}$, and depends on the direction $\vec{v}$ of the
charged particles  and the geomagnetic field $\vec{B}$. In
contrast to the charge excess mechanism, the geomagnetic emission of
electrons and positrons add up since they have opposite charges but
their trajectories curve in opposite directions (see
Section~\ref{sec:mechanisms}). In the top panel of Fig.~\ref{fig:polvertical} 
we show schematically the projection of the polarization on the ground
induced by each of the emission mechanisms for a vertical
shower~\cite{REAS3}. A horizontal magnetic field pointing north (N) is assumed.

The interplay between the different polarizations makes the projection of the
net field on ground asymmetric with respect to the shower core~\cite{polinterference}. This can be easily understood from the sketch at the top of Fig.~\ref{fig:polvertical}.
Since the Askaryan and geomagnetic polarizations point in the same direction
for observers East of the core, and in opposite directions for observers to
the west, we
expect the EW component of the electric field ($E_{EW}$) to be larger in an antenna eastwards of the shower
core than in an antenna West of the core, at the same distance.  We also expect $E_{EW}$  to be of the same order northwards and southwards from
the shower core. 
Another prediction is that the NS component of the electric field ($E_{NS}$) should be close to zero
for observers along the East-West direction and largest for observers along
the North-South one \cite{REAS3}. As discussed in \cite{polinterference}, as the
distance $r$ from the antenna to the shower axis increases, the relative
contribution to the electric field of the geomagnetic mechanism decreases and
as a consequence the Askaryan mechanism accounts for a larger fraction of the
emission. The pattern of the electric field at ground level is then expected
to recover a symmetric behavior with respect to the shower core for very large
distances, when the Askaryan contribution dominates.

To investigate these expectations we have obtained the $E_{EW}$ and $E_{NS}$ 
components of the field, at a frequency of $60$~MHz, from ZHAireS simulations of vertical proton showers with energy $10^{17}$ eV. 
These components are shown in Fig.~\ref{fig:polvertical}. In the left panel 
we show $E_{NS}$ as a function of the position on the ground, 
while in the right panel we show the corresponding $E_{EW}$.  
One can see the expected EW asymmetry in the $E_{EW}$ component, with larger fields 
to the east of the core, while it is approximately the same North and South of the
shower core. The $E_{NS}$ component (left panel) is mainly due 
to the Askaryan mechanism, and hence it is largest along the NS direction, while 
it gradually decreases as the observer moves to the EW direction because
neither the Askaryan nor the geomagnetic mechanism induce a significant
$E_{NS}$ component along observers in the EW direction (see sketch in Fig.~\ref{fig:polvertical}). 

Comparing $E_{NS}$ and $E_{EW}$ for observers along the NS axis, one can clearly see that at 
distances relatively close 
to the shower core ($r\lesssim 150$~m), the NS component of the
field is a factor $\sim 4$ smaller than the EW component,  
because the geomagnetic mechanism dominates the emission close 
to the core, while this factor tends to diminish at larger distances (see also
Fig.~\ref{fig:ldfplpot}). 

\subsection{ $B=0$ vs $B\neq0$: Separation of the geomagnetic and Askaryan components.}

\begin{figure}
\begin{center}
\scalebox{0.35}{
\includegraphics{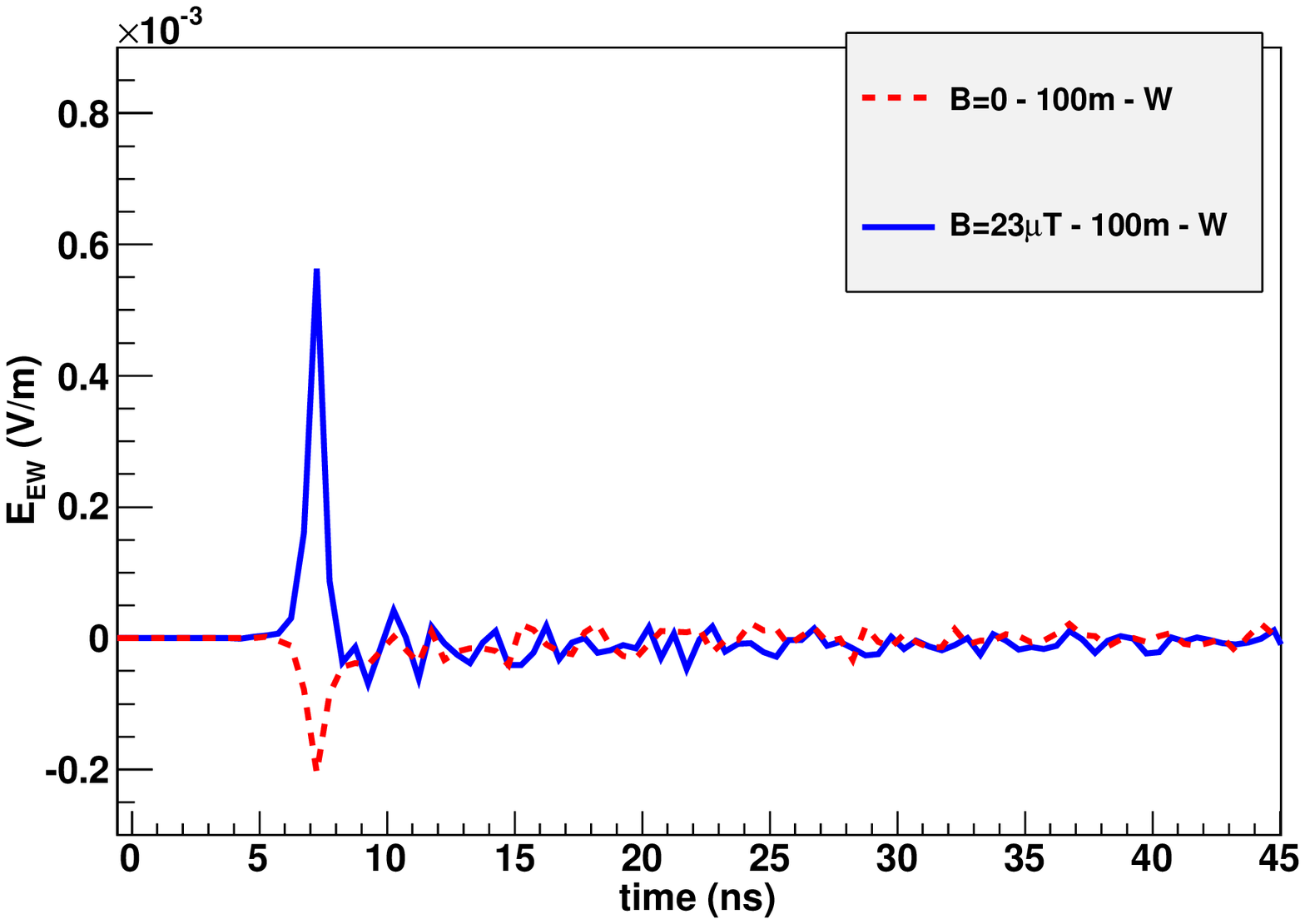}
\includegraphics{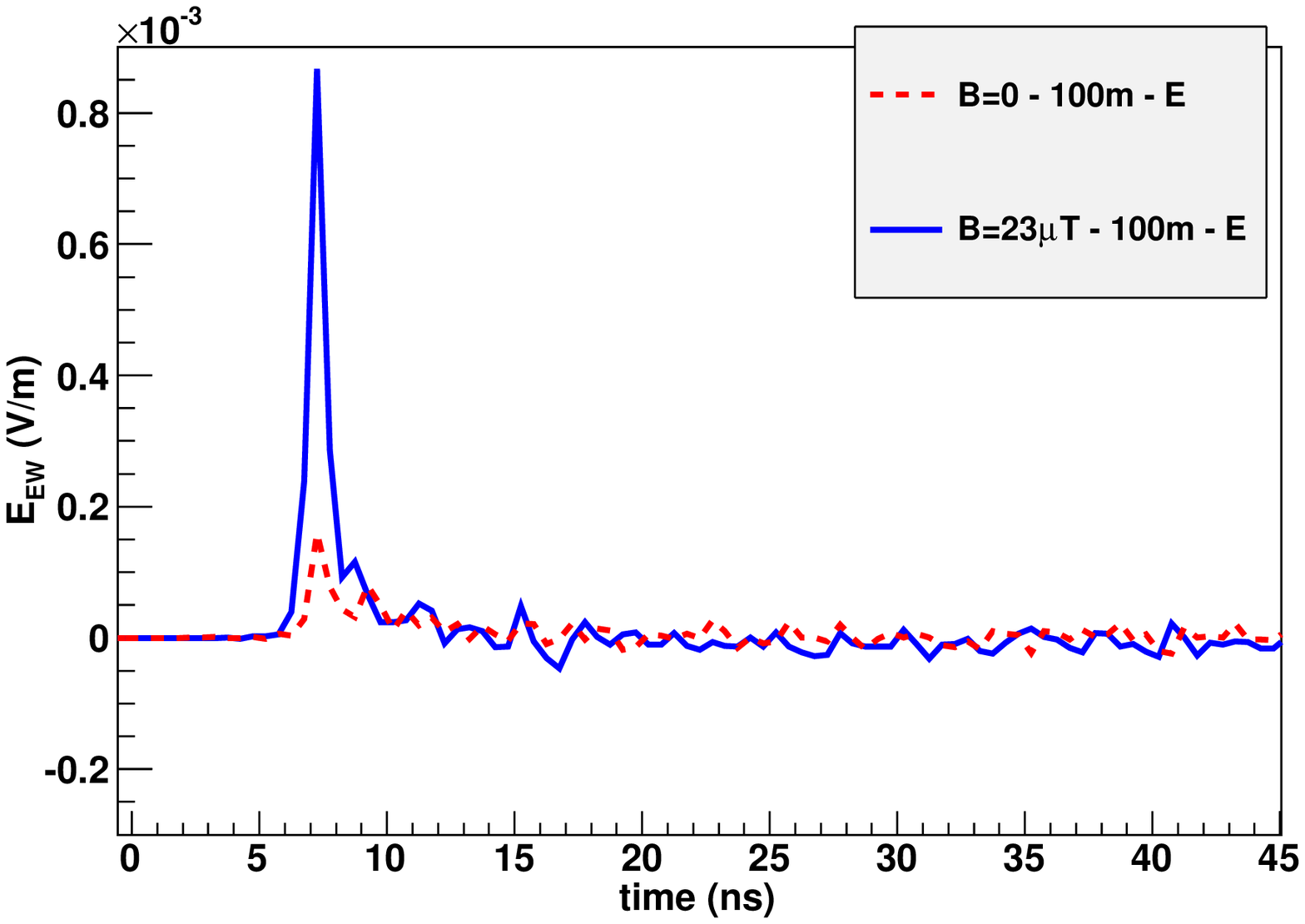}}
\scalebox{0.35}{
\includegraphics{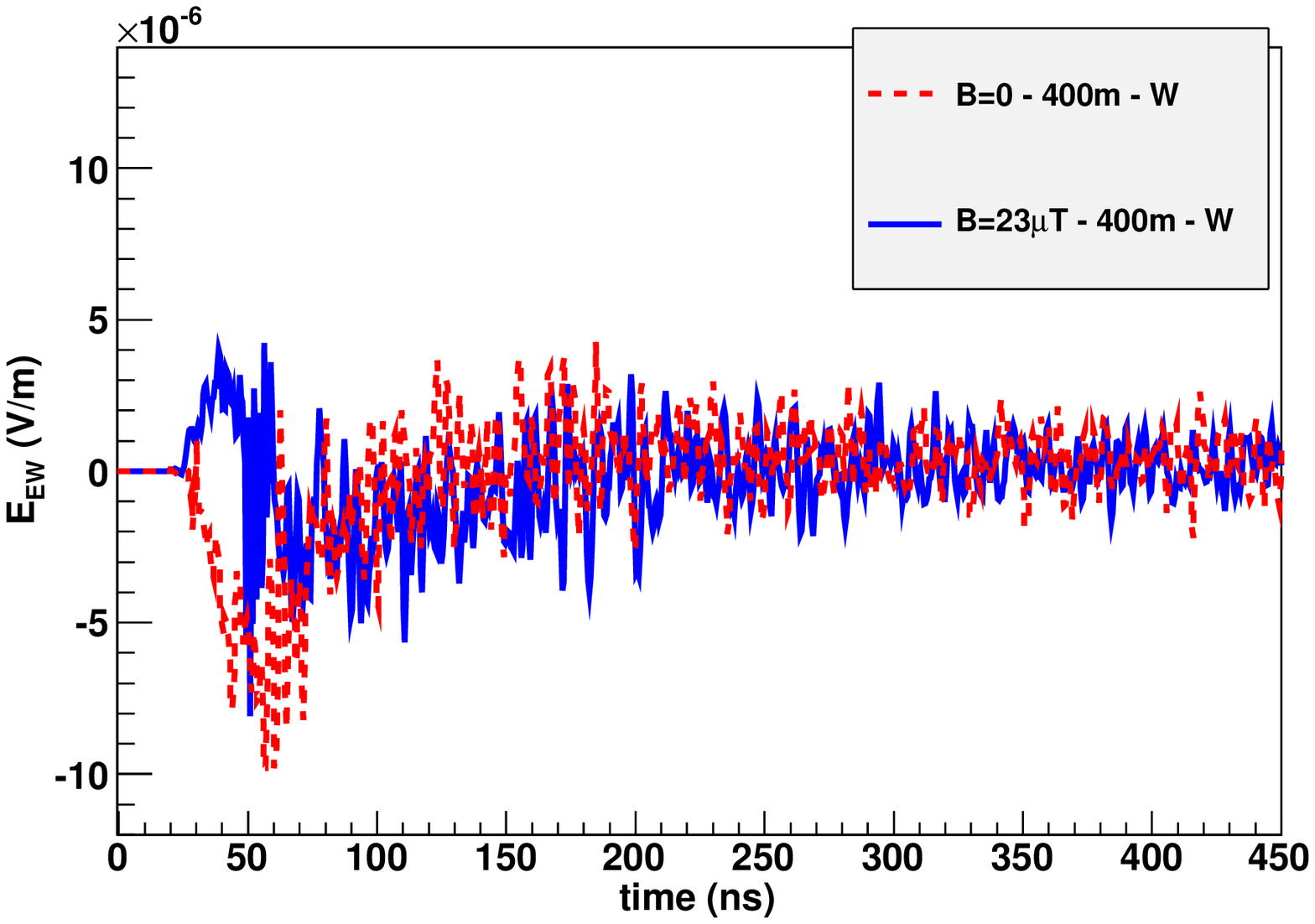}
\includegraphics{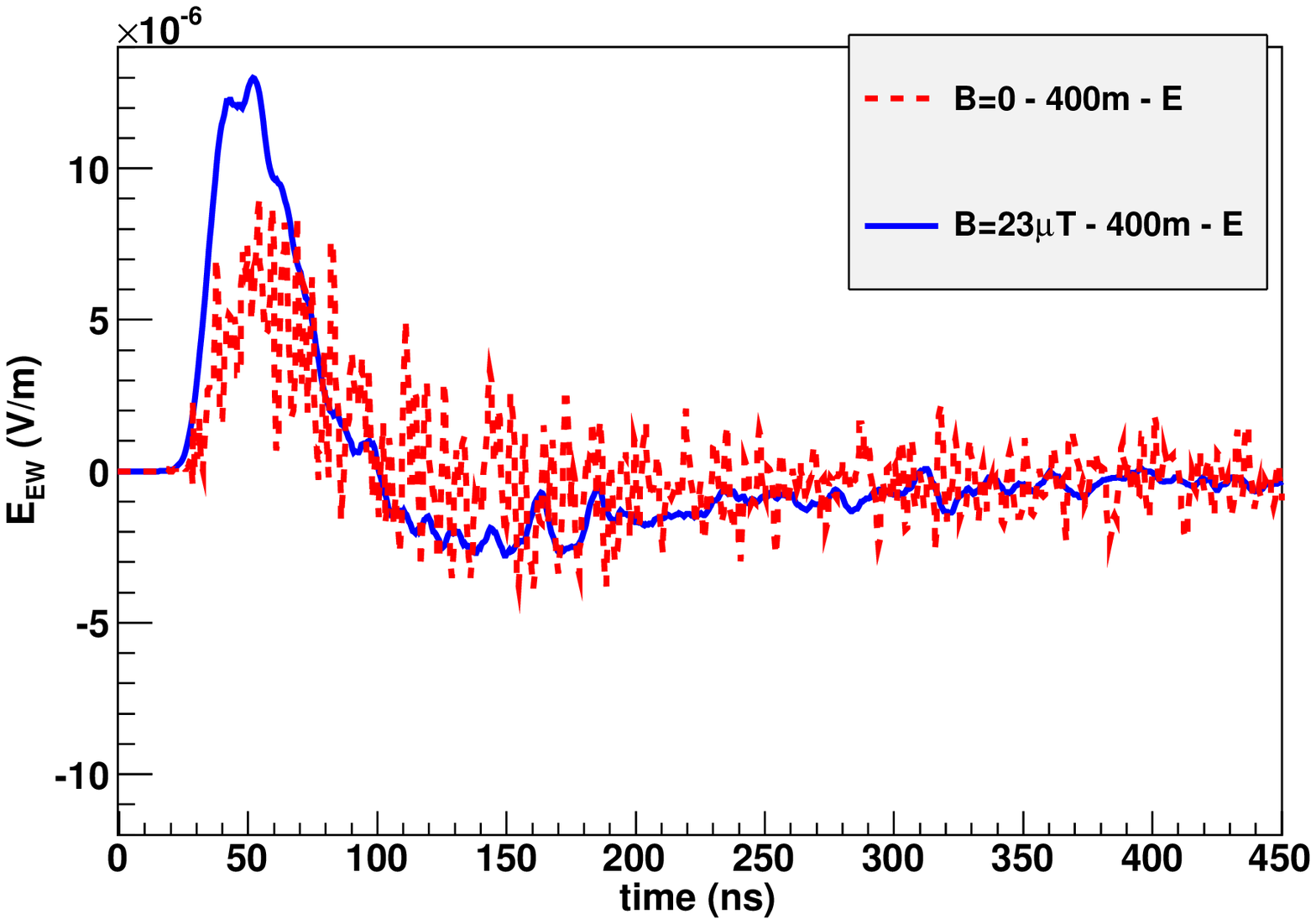}}
\caption{Comparison between the EW component of the field calculated with and without the
  geomagnetic field at a $100$~m (top) and $400$~m (bottom) West (left) and East
  (right) of the core.}
\label{fig:comp-B-noB}
\end{center}
\end{figure}

In order to disentangle the geomagnetic and Askaryan components of the radio
emission, we also simulated showers turning off the geomagnetic field. As
discussed in section \ref{sec:polarization}, there is an EW asymmetry in the
signal strength due to an interference effect between the  polarizations of
the different emission mechanisms, making the signal larger to the East. In
fig. \ref{fig:comp-B-noB} we compare the EW components (positive to the East)
at $100$~m (top) and $400$~m (bottom) West (left) and East (right) of the core,
obtained with simulations with and without the geomagnetic field. The
simulation with the magnetic field on shows the net field due to both emission
mechanisms, while  the simulations without the magnetic field has only the
Askaryan  component. One can see that, as expected, the polarization of the
Askaryan component changes direction between antennas East and West of the
core (dashed red lines of fig. \ref{fig:comp-B-noB}), and this causes the
difference in the peak height of the net field (solid blue lines). Also, the
height of the pure Askaryan peak (dashed red) is roughly half the difference between
the E and W peaks of the net field (solid blue), as expected~\cite{REAS3} (see
also the sketch in Fig.~\ref{fig:polvertical}) . At larger
distances (e.g. at $400$~m, at the bottom of fig.\ref{fig:comp-B-noB}) , the
ratio between the Askaryan and geomagnetic components gets higher. At larger distances, the Askaryan mechanism starts to dominate the emission.

To further investigate the polarization of the emission we use a parameter ${\cal R}_p$,  as defined in~\cite{Auger_Harm}, which is sensitive
to the polarization of the electric field.
In the case of a horizontal (parallel to ground) magnetic field 
pointing north, the polarization vector of the geomagnetic
contribution to the electric field points East, and ${\cal R}_p$ is given by: 

\begin{equation}
{\cal R}_p=\frac{\sum_t{E_{EW} \cdot E_{NS}}}{\sum_t{(E_{EW}^2+E_{NS}^2)}}\;\;,
\label{eq:R}
\end{equation}
where the sum runs over all bins in time having a non-zero electric field
at the antenna.

\begin{figure}
\begin{center}
\scalebox{0.7}{
\includegraphics{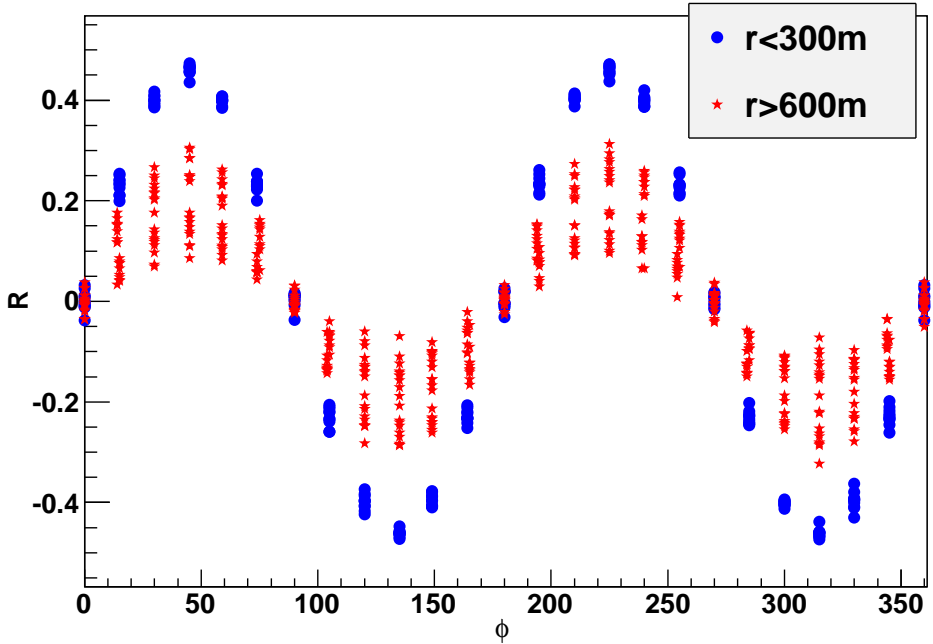}}
\caption{
ZHAireS code predictions for the polarization sensitive
parameter ${\cal R}_p$ (as defined in Eq.~(\ref{eq:R})), as a function of 
the azimuthal angle $\phi$ (see top of Fig.~\ref{fig:polvertical}) for 10 vertical proton showers of 
energy $10^{17}$ eV. The magnetic field was switched off in the simulations. 
The ${\cal R}_p$ parameter is plotted for antennas close to the shower
core (blue circles: $r<300$~m), and farther from it (red stars: $r>600$~m).}
\label{fig:Rplot_Boff}
\end{center}
\end{figure}

From the sketch in the top of Fig.~\ref{fig:polvertical} we expect 
${\cal R}_p$ to vary as a function of the azimuthal angle $\phi$ on the ground, 
defined so that $\phi=0$ for an antenna in the East direction and 
$\phi=90^\circ$ for an antenna at the North. ${\cal R}_p=0$ when 
either $E_{EW}$ or $E_{NS}$ equal zero. We then expect that 
if the polarization were only due to the geomagnetic contribution (the
unphysical case of no charge excess in the shower), the NS polarization would
always be zero and so would ${\cal R}_p$.
If at a certain distance $r$ to the shower core the Askaryan and geomagnetic
contributions are important, then
${\cal R}_p\sim0$ at $\phi=0^\circ$ and $\phi=180^\circ$, 
because $E_{NS}\sim0$ along those two directions~\cite{Auger_Harm}.
However, in contrast to what was expected in~\cite{Auger_Harm}, if the geomagnetic component is absent then we clearly expect ${\cal R}_p\sim0$ at $\phi=0^\circ,~90^\circ,~180^\circ$
and $270^\circ$, because either $E_{EW}\sim 0$ or $E_{NS}\sim 0$ 
along those directions. This means that in the absence of 
a magnetic field, ${\cal R}_p$  should exhibit 
a $180^\circ$ periodicity in azimuthal angle $\phi$. We checked this in 
our ZHAireS simulations by switching off the magnetic field. The result
is plotted in Fig.~\ref{fig:Rplot_Boff} where one can see that 
${\cal R}_p$ exhibits a periodicity of $180^\circ$ in $\phi$, 
regardless of the distance to the observer.

\begin{figure}
\begin{center}
\scalebox{0.7}{
\includegraphics{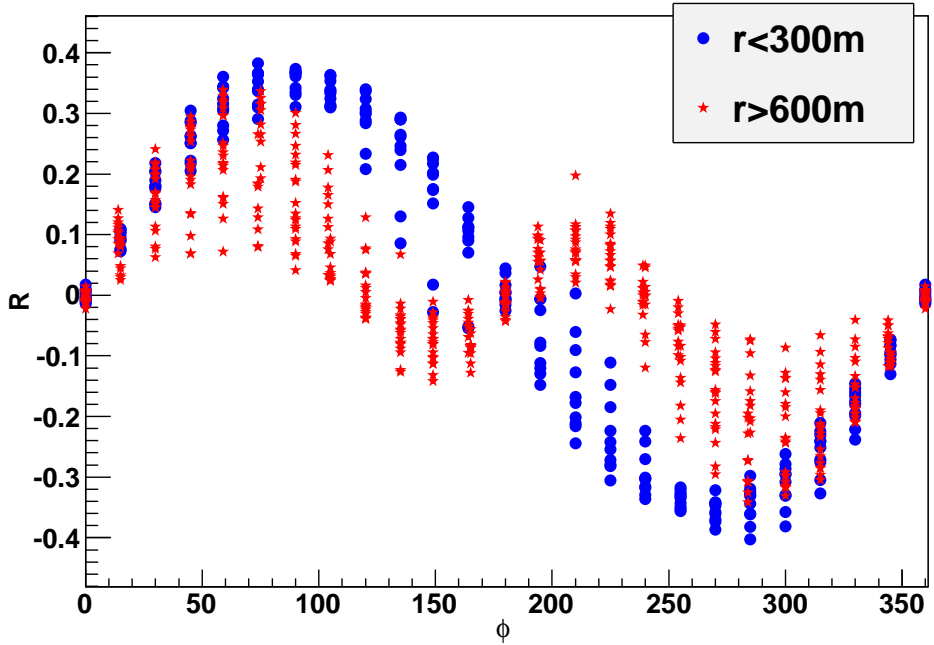}} 
\caption{ZHAireS code predictions for the polarization sensitive
parameter ${\cal R}_p$ (as defined in Eq.~(\ref{eq:R})), as a function of 
the azimuthal angle $\phi$ for 10 vertical proton showers of 
energy $10^{17}$ eV, with a horizontal magnetic field pointing north.
The ${\cal R}_p$ parameter is plotted for antennas close to the shower
core (blue circles: $r<300$~m)  and farther from it (red stars: $r>600$~m).}
\label{fig:Rplot}
\end{center}
\end{figure}

In contrast, in Fig. \ref{fig:Rplot} we show ${\cal R}_p$ as a function of $\phi$ 
as obtained in ZHAireS simulations of 10 vertical proton showers of 
energy $10^{17}$ eV, but in this case with the geomagnetic field on.
One can see that for distances closer to the core ($r<300$~m), the ${\cal R}_p$ 
parameter reaches zero only at  $\phi=0^\circ$ and $\phi=180^\circ$, as expected
when both the geomagnetic and Askaryan components are important, while at larger distances from the
core, the period of ${\cal R}_p$ vs $\phi$ changes from $360^\circ$ to $\sim
180^\circ$. This further confirms that at large distances to the core
($r\gtrsim 600$ m),
the Askaryan mechanism dominates the full bandwidth emission. A similar behavior was
reported in~\cite{Auger_Harm} for REAS3 simulations convoluted with the
detector response, but this periodicity change in ${\cal R}_p$ was interpreted as a signature of a dipole polarized field instead of a dominant
charge excess contribution. In the same paper~\cite{Auger_Harm}, MGMR simulations did not exhibit this behavior.

If we assume a perfect shower symmetry in $\phi$, the modulus $E^{\rm Ch}$ of
the electric field due to Askaryan mechanism would be the same for antennas at
the same distance from the core, and the EW and NS components could be written
as:

\begin{equation}
E^{\rm Ch}_{EW}=-E^{\rm Ch}\cos{\phi},\;\;\;
E^{\rm Ch}_{NS}=-E^{\rm Ch}\sin{\phi}
\label{eq:askaryan}
\end{equation}  

If we further assume no time dependence of the polarization, and take the
numerator of Eq.~(\ref{eq:R}):
\begin{equation}
{\cal R}_p^{\rm Ch}\propto
{(E^{\rm Ch})}^2\sin{2\phi}\;\;,
\label{eq:Raskaryan}
\end{equation}  
\noindent which leads to ${\cal R}_p^{\rm Ch}$ with a period of 180$^{\circ}$ in $\phi$, as shown in Fig.~\ref{fig:Rplot_Boff}.

The field with both the Askaryan $E^{\rm Ch}$ and geomagnetic $E^{\rm geo}$
contributions can be written as (see top of Fig.~\ref{fig:polvertical}):
\begin{equation}
E_{EW}=-(E^{\rm geo}+E^{Ch}\cos{\phi}),\;\;\;E_{NS}=-E^{\rm Ch}\sin{\phi}
\label{eq:netfield}
\end{equation}  
\noindent leading to:
\begin{equation}
{\cal R}_p \propto E^{\rm geo}~E^{\rm Ch}\sin{\phi} + {(E^{\rm Ch})}^2 \sin{2\phi}/2
\label{eq:netfieldR}
\end{equation}  
The equation above predicts that  
when the Askaryan and geomagnetic mechanisms compete, ${\cal R}_p$ is
proportional to a rather 
complicated function of $\phi$ whose shape depends on the relative values 
of $E^{\rm Ch}$ and $E^{\rm geo}$.

\begin{figure}
\begin{center}
\scalebox{0.35}{
\includegraphics{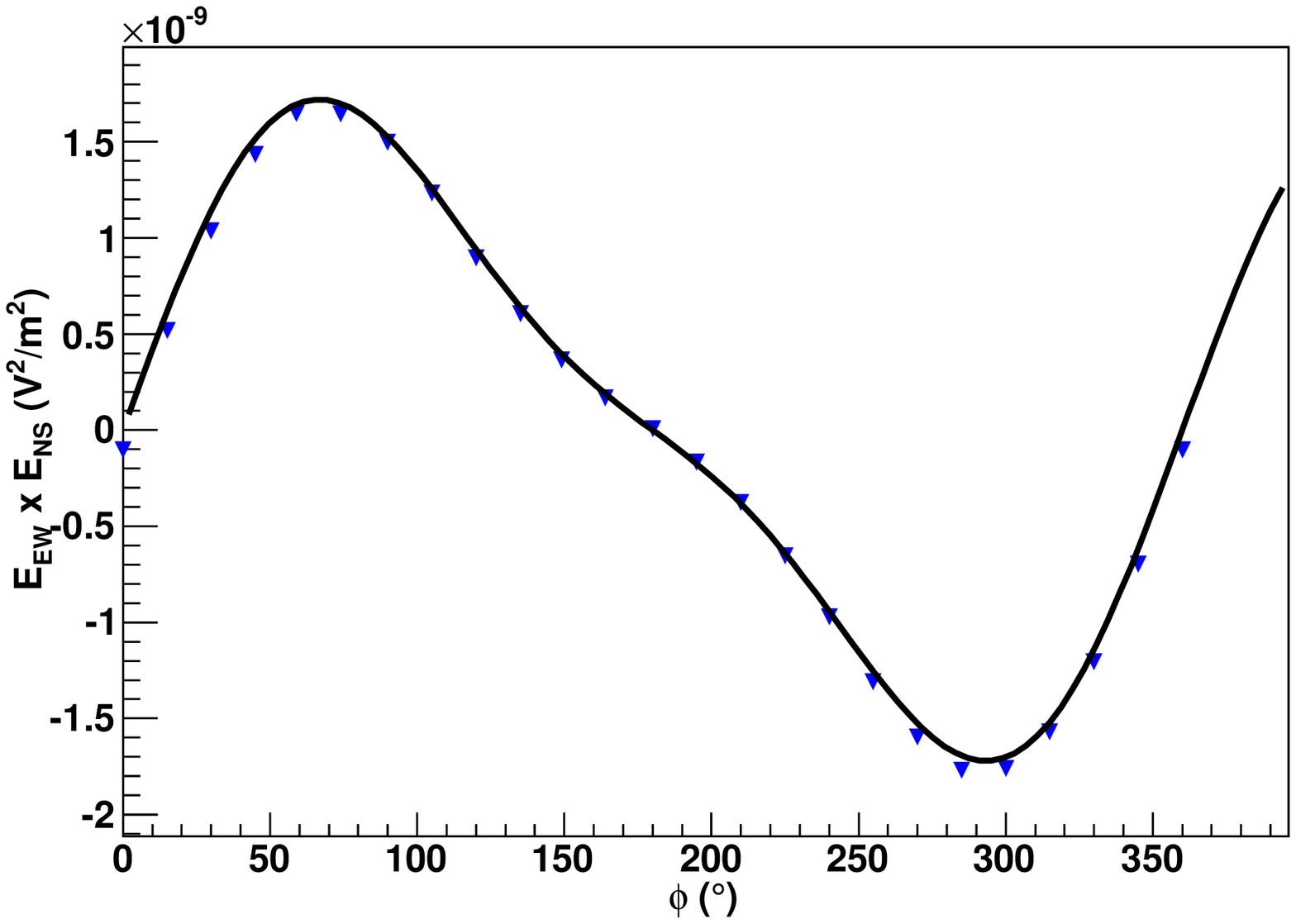} \includegraphics{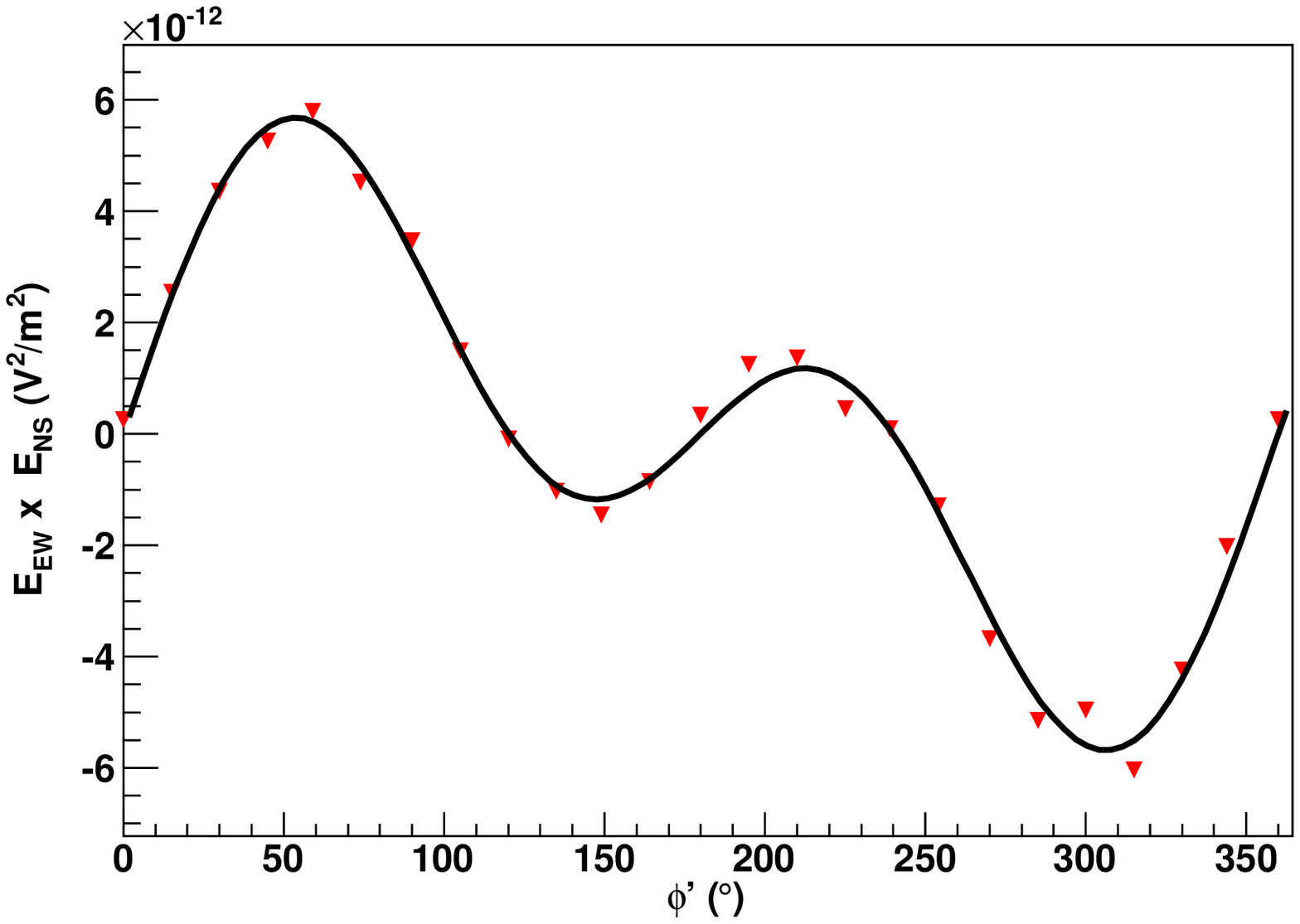}}
\caption{Points: The numerator of ${\cal R}_p$ in Eq.(\ref{eq:R}) 
as a function of $\phi$ for a vertical proton shower of $10^{17}$ eV 
simulated with ZHAireS, for antennas at $r=200$~m (left) and $r=1200$~m (right) from
the core. Solid lines: Fit of Eq.~(\ref{eq:netfieldR}) (obtained using a simple model of the polarization - see text for more details) to the simulated electric fields.}
\label{fig:numplot}
\end{center}
\end{figure}

In Fig.~\ref{fig:numplot} we show the numerator of ${\cal R}_p$, 
given by the product $E_{EW} \cdot E_{NS}$, as a function of $\phi$ 
obtained in ZHAireS simulations of a vertical proton shower of $10^{17}$ eV,
for antennas at $r=200$~m (left) and $r=1200$~m (right) from the shower core. 
We then fitted  Eq.~(\ref{eq:netfieldR}) to these simulations 
with only $E^{\rm Ch}$ and $E^{\rm geo}$ as free parameters. The fit based
on the simple model above reproduces
remarkably well the behavior with $\phi$ obtained in the simulations, with a
ratio $E^{\rm geo}/E^{\rm Ch}\sim 1.74$ at $r=200$~m, and $E^{\rm geo}/E^{\rm
  Ch}\sim 0.51$ at $r=1200$~m, decreasing with $r$ as the Askaryan component
is expected to dominate at large distances to the core. It is interesting to
note that in the ZHAireS simulation for antennas south of the core, the ratio
between the peaks of $E_{EW}$ (expected to be purely geomagnetic) and $E_{NS}$
(expected to be purely Askaryan) decreases from $2.32$ at $r=200$~m to $0.57$
at $r=1200$~m, for this same shower. It is also interesting to see in
Fig.~\ref{fig:numplot} how the parameter ${\cal R}_p$ gradually changes from
having a $360^\circ$ periodicity in $\phi$ to a $180^\circ$ periodicity as the
distance to the shower axis increases. We believe that similar, but more
refined analysis methods could be derived to separate the contributions of the
different emission mechanisms to the electric field.

\subsection{Polarization of the electric field in non-vertical showers}

\begin{figure}
\begin{center}
\scalebox{0.6}{
\includegraphics{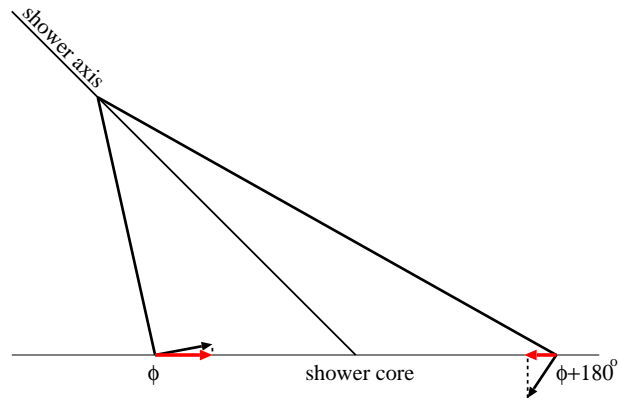}}
\caption{Sketch of the electric field in non vertical showers and its
  projection onto the ground plane.}
\label{fig:nonverticalpol}
\end{center}
\end{figure}
\begin{figure}
\begin{center}
\scalebox{0.6}{
\includegraphics{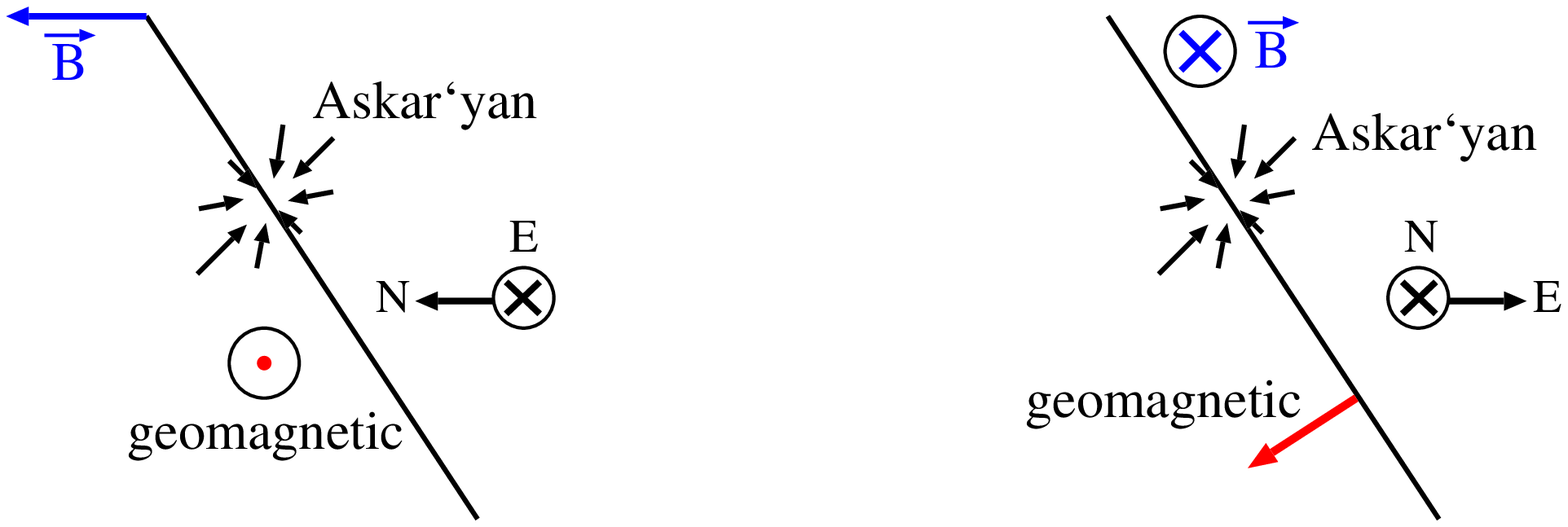}}
\scalebox{0.35}{
\includegraphics{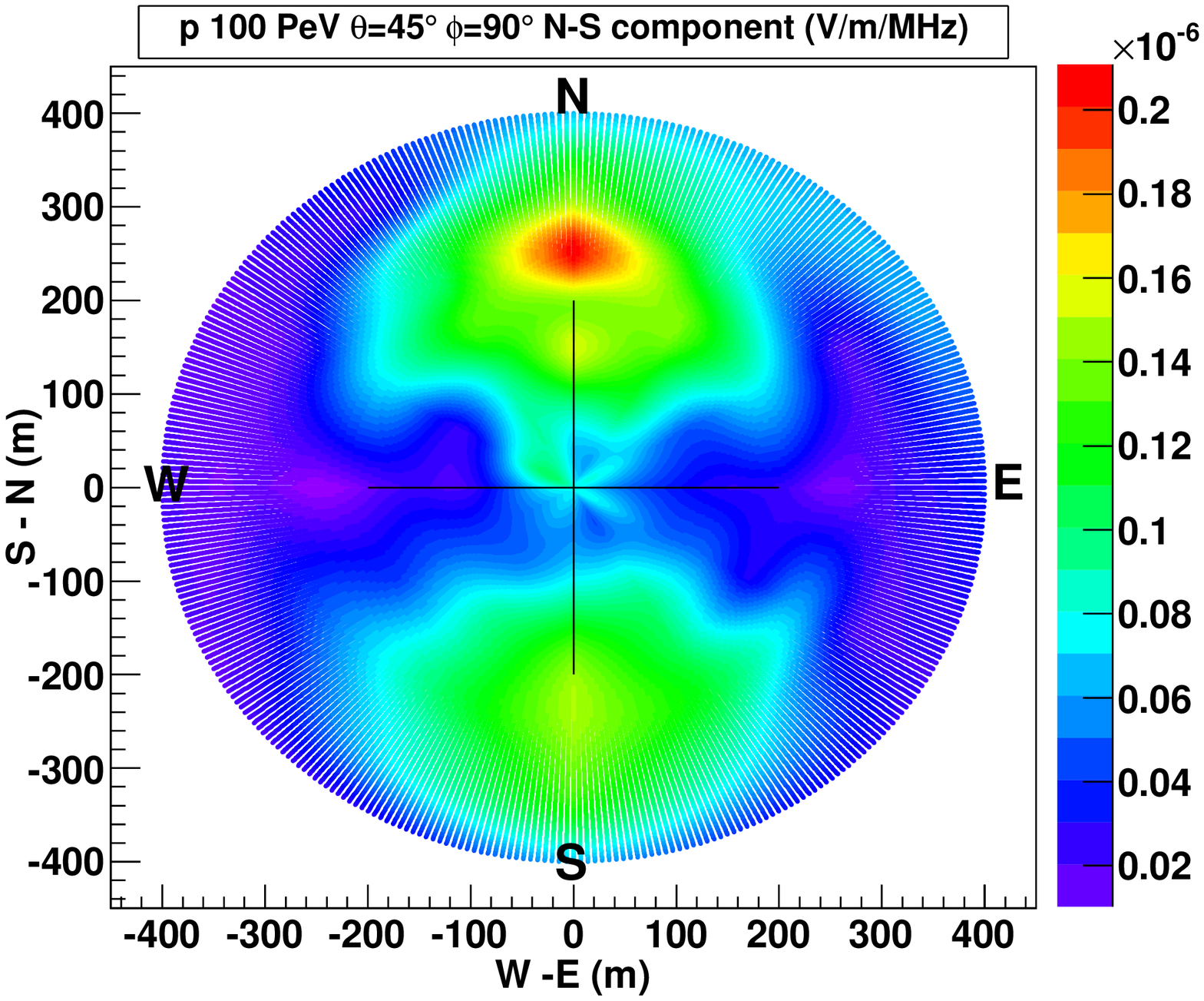}\includegraphics{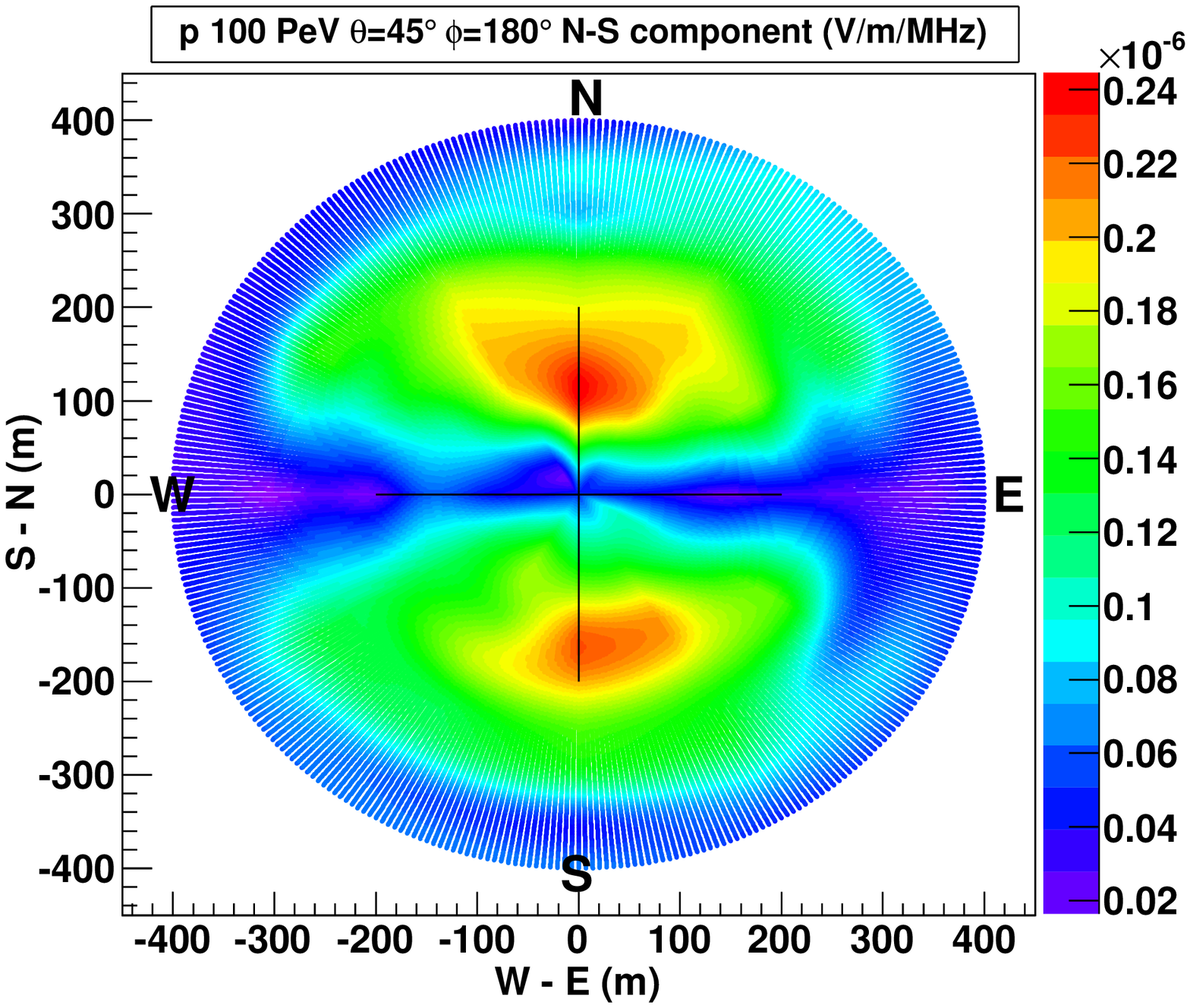}}
\scalebox{0.35}{
\includegraphics{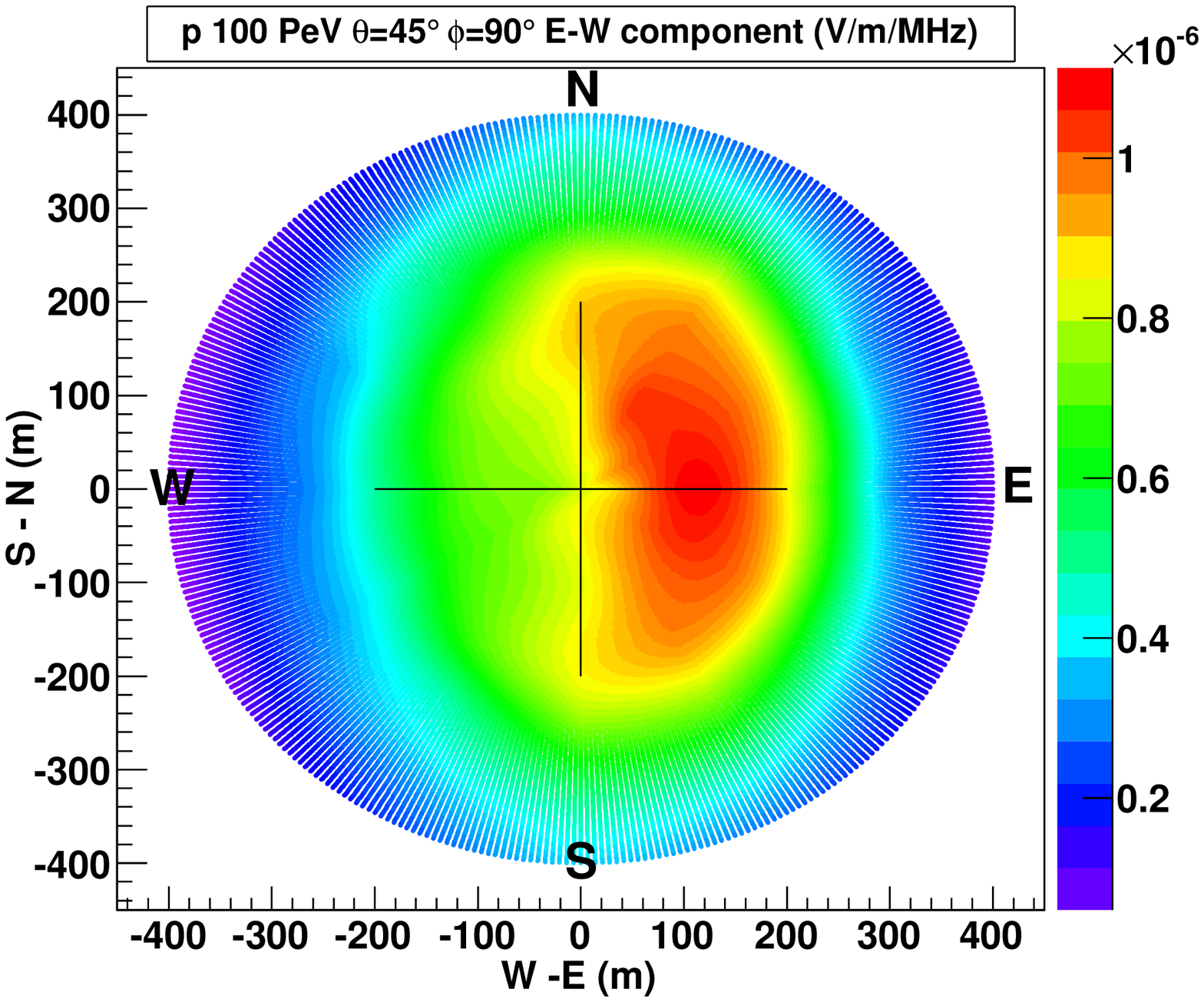}\includegraphics{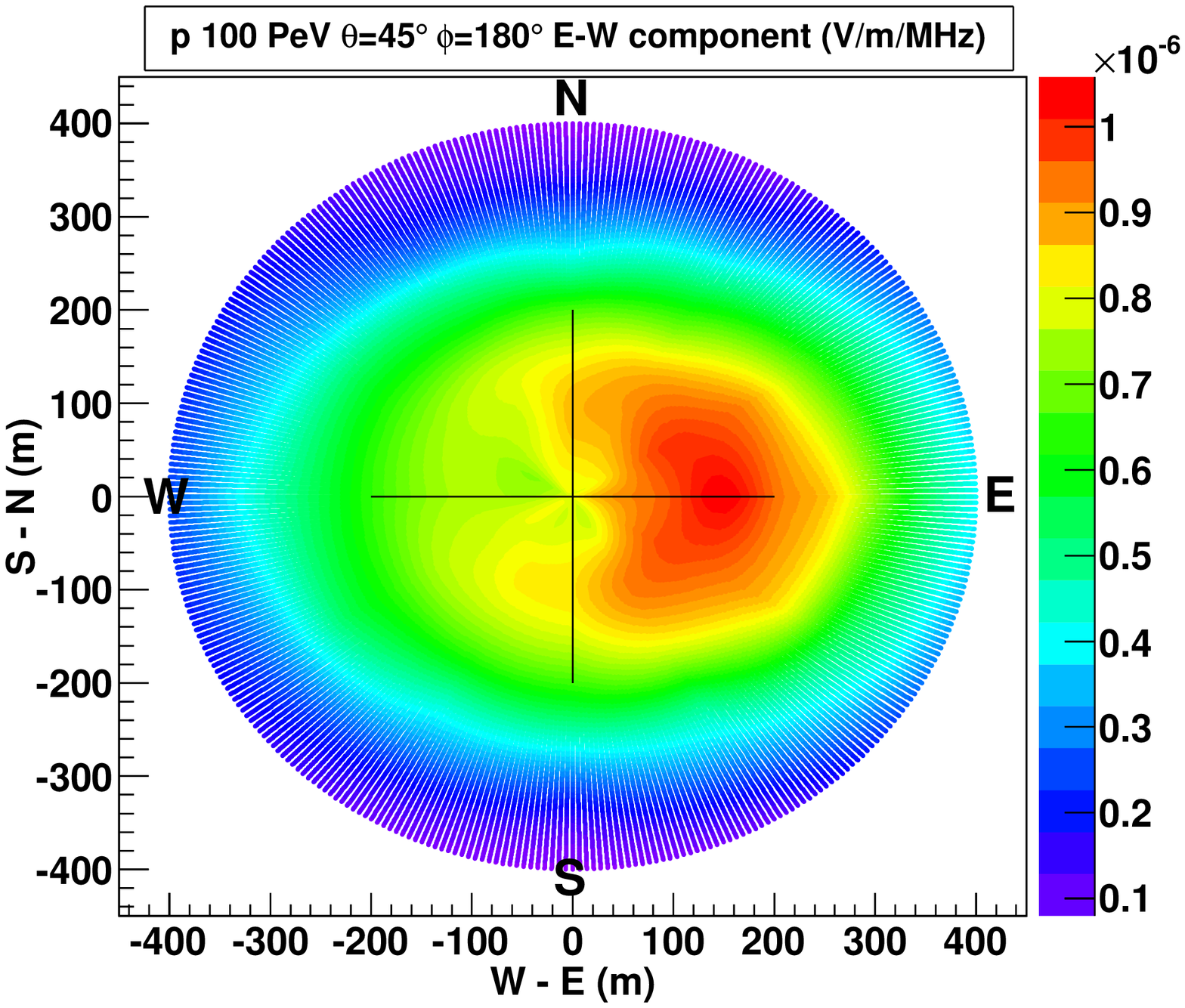}}
\scalebox{0.35}{
\includegraphics{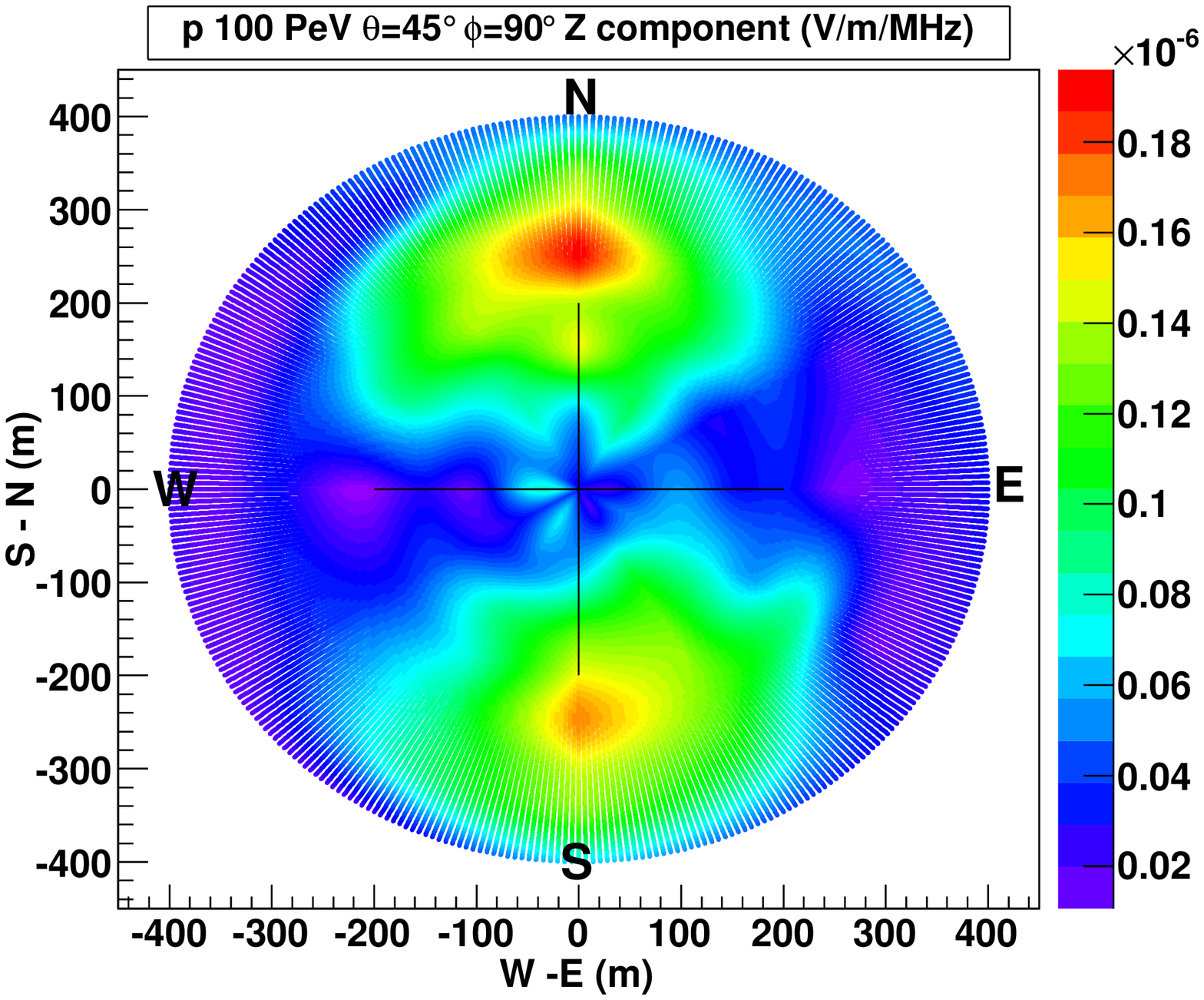}\includegraphics{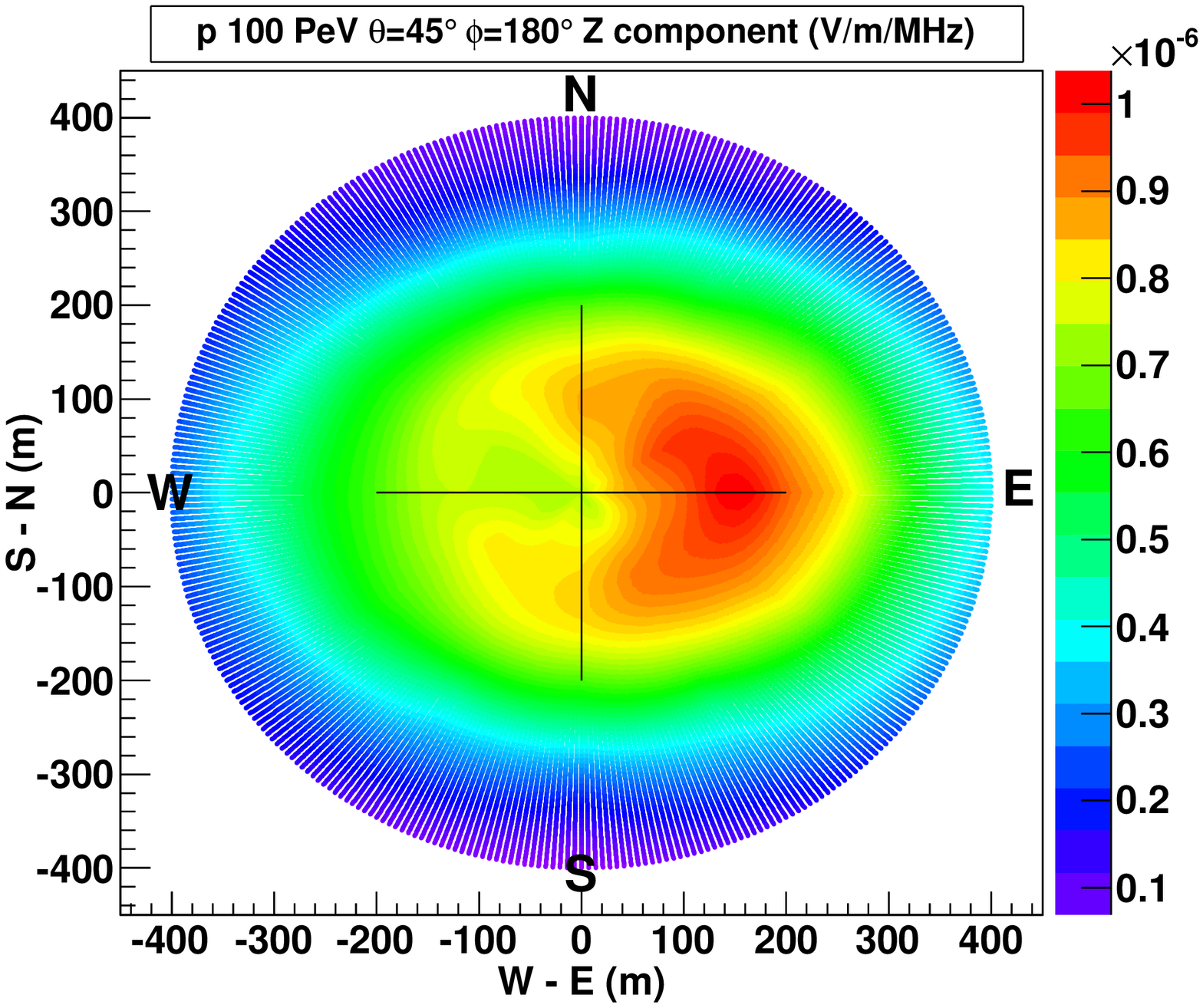}}
\caption{Results of full ZHAireS
simulations of the N-S (top), E-W (middle) and
Z (bottom) components of the electric field at 60MHz for 100PeV proton
shower with $\theta=45^\circ$ coming from the north (left) and the west
(right). See text for more details.}
\label{fig:pol-45}
\end{center}
\end{figure}

In the case of non-vertical showers, the projection on the ground of the polarization of
both, the geomagnetic and Askaryan components will depend on the azimuthal
angle $\phi$ of the antenna position on the ground in relation to the core,
as can be seen in schematic form in
Fig. \ref{fig:nonverticalpol}. Furthermore, the direction of the shower axis
with respect to the geomagnetic field will change the direction of the
Lorentz force acting on the charged particles. In vertical showers the force
is always almost parallel to the ground, but in non-vertical showers
the Lorentz force will only be horizontal if the plane defined by $\vec{B}$ and the shower axis is
perpendicular to the ground
(e.g. Fig. \ref{fig:pol-45} top left). In most geometries there will be a vertical component to
the Lorentz force (e.g. Fig. \ref{fig:pol-45} top right). This will cause a dependence of the
polarization and the signal asymmetries on the azimuthal angle of the shower, as can be
seen in Fig. \ref{fig:pol-45}, which shows the results of full ZHAireS
simulations of the N-S (top), E-W (middle) and
Z (bottom) components of the electric field at $60$~MHz for $100$ PeV proton
shower with $\theta=45^\circ$ coming from the north (left) and the west
(right), corresponding to the geometries shown schematically on the top left and top
right of Fig. \ref{fig:pol-45}, respectively. Note that different scales were
used for the field in the different panels. A horizontal magnetic field
pointing north was used in the simulations. For the shower coming from the north
(left), one can see that the main asymmetry is in the E-W component (left-middle), similar to
the vertical shower case, while the N-S and Z components are smaller
and very similar to each other, because the  dominant geomagnetic
contribution is horizontal. On the other hand, in the case of the shower coming from the West
(right), there is a large asymmetry to the East on both, the E-W and Z
components, since in this particular geometry the dominant geomagnetic
contribution makes an angle of $\sim45^\circ$ with the horizontal, and thus should
have very similar E-W and Z components.  

This dependence of the polarization on the azimuthal angle of the shower
direction is relevant for studies trying to disentangle the contributions of
the emission mechanisms using polarization. Since the inclination of the
polarization vector of the dominant geomagnetic emission with respect to
the horizontal plane changes with the shower azimuthal angle, it may be
important to also measure the vertical component of the net
electric field, since for showers with $\theta>45^\circ$, the vertical
component of the field can be even larger than the horizontal ones.

\section{Conclusions}
\label{sec:conclusion}

In this work we present predictions for the radio pulse emitted by extensive 
air showers. Our results are obtained using 
the ZHAireS Monte Carlo, an AIRES-based code that takes into account the full 
complexity of ultra-high energy cosmic-ray induced shower development
in the atmosphere, 
and allows the calculation of the electric field in both the time and frequency domains
based on the algorithms developed in \cite{ZHS92,ARZ10}. Although our
approach does not presuppose any a priori emission mechanism, our
results confirm that the emission at radio frequencies  
can be understood as the superposition of radiation from the charge separation induced
by the magnetic field of the Earth (geomagnetic effect), and that coming from the net 
excess negative charge evolving as the shower develops in the atmosphere (Askaryan
effect).

We have pointed out the relevance of the refractive index in the time structure and 
intensity of the radio pulses, especially at short distances to the shower
axis. Another interesting work~\cite{scholtenprl}, developed independently and
in parallel with ours, deals with similar issues. The refractive index determines the angular distribution of the radiation at 
the emission point, as well as its propagation through the atmosphere
and non trivial relativistic effects arise due to the refractive index $n>1$. 
We have developed a simple 1-dimensional model to address the
characteristics of the radio pulse, which qualitatively allows us to
interpret the pulse height, width and its dependence on the refractive
index and the distance from antenna to shower core \cite{scholtenprl}. The intensity of the radio pulses 
is typically highest for those observers that see a region of the shower containing 
a large number of particles (such as shower maximum) with an angle close to the 
Cherenkov angle. There is a non-trivial interplay between the distance 
from the emission point in the shower to the observation point, the angle between 
the particle direction and the observer, and the number of particles in the region of the 
emission, whose effects can only be accurately determined with Monte Carlo simulations
such as those developed in this work. These key elements determine for instance that 
far from the shower axis inclined showers typically induce a more intense radiation 
than vertical ones \cite{Gousset_inclined} as we show with our ZHAireS simulations. 

We have shown that the frequency at which the emission spectrum is maximum, which
is in the range $\sim3$-$30$~MHz for $0<r<400$ m, decreases as the distance
$r$ from the antenna to the shower core increases (for $r>100$~m see also
\cite{Scholten_MGMR,REAS3,compREAS3MGMR}). So most experiments, which
typically are only sensitive to frequencies above $30$~MHz, in fact only measure the incoherent part of the spectrum, even very close to the shower core. We also observed that the
signal at higher frequencies decreases more rapidly with $r$ than at the fully coherent 1 MHz
range, where we observed a maximum in the emission at around $r\sim 100$ m,
similar to the one reported in \cite{scholtenprl}. Our results also suggest
that this maximum is dependant on frequency and on observer direction w.r.t. the shower core. We have also explored the
polarization properties of the radiation, confirming 
the expectation that the radiation is mainly polarized 
in the opposite direction to the Lorentz force induced by the magnetic field of the Earth~\cite{kahnlerchegeo,lerchenature,codalemavxb,scholtenarena2010}. 
Far from the shower core, our simulations show that the polarization is
compatible with the presence of a significant amount of radiation due to the Askaryan effect, as in~\cite{REAS3}. We have also
shown, using a very simple model, that as the Askaryan component of
the emission becomes dominant over the geomagnetic one at larger distances to
the core, the parameter ${\cal R}_p$ changes its periodicity on the azimuthal
angle of the observer. When both, the geomagnetic and Askaryan components are
significant, the periodicity is $360^\circ$, but when the emission is
dominated by the Askaryan effect, a $180^\circ$ periodicity can be clearly appreciated. In inclined showers we have stressed the role played by the
relative orientation of the shower axis and the magnetic field on the
polarization. A significant vertical component of the electric field arises in
inclined showers which calls for detection instruments able to observe it.  

\section{Acknowledgments}
We thank Xunta de Galicia (INCITE09 206 336 PR) and
Conseller\'\i a de Educaci\'on (Grupos de Referencia Competitivos --
Consolider Xunta de Galicia 2006/51); Ministerio de Ciencia e Innovaci\'on
(FPA 2007-65114, FPA 2008-01177 and Consolider CPAN - Ingenio 2010)
and Feder Funds, Spain. We thank CESGA (Centro de SuperComputaci\'on
de Galicia) for computing resources and assistance.
We thank M. Tueros, J. Bray, T. Huege, C.W. James and A. Romero-Wolf
for helpful discussions.

\end{document}